\documentclass{emulateapj}

\slugcomment{Accepted for Publication in The Astrophysical Journal}

\shorttitle{Star-Forming Galaxy Unveiled with ALMA and HST}
\shortauthors{Ouchi et al.}

\begin{document}

\title{
An Intensely Star-Forming Galaxy at $z\sim 7$ 
with Low Dust and Metal Content Revealed by Deep ALMA and {\sl HST} Observations
}

\author{
Masami Ouchi        \altaffilmark{1,2},
Richard Ellis       \altaffilmark{3},
Yoshiaki Ono        \altaffilmark{1},
Kouichiro Nakanishi \altaffilmark{4,5},
Kotaro Kohno        \altaffilmark{6,7},\\
Rieko Momose        \altaffilmark{1},
Yasutaka Kurono       \altaffilmark{5},
M. L. N. Ashby \altaffilmark{8},
Kazuhiro Shimasaku  \altaffilmark{7,9},\\
S. P. Willner    \altaffilmark{8},
G. G. Fazio   \altaffilmark{8},
Yoichi Tamura       \altaffilmark{6},
and
Daisuke Iono        \altaffilmark{10},
}

\altaffiltext{1}{Institute for Cosmic Ray Research, The University of Tokyo, 5-1-5 Kashiwanoha, Kashiwa, Chiba 277-8582, Japan; ouchims@icrr.u-tokyo.ac.jp}
\altaffiltext{2}{Kavli Institute for the Physics and Mathematics of the Universe (WPI), The University of Tokyo, 5-1-5 Kashiwanoha, Kashiwa, Chiba 277-8583, Japan}
\altaffiltext{3}{Department of Astrophysics, California Institute of Technology, MS 249-17, Pasadena, CA 91125, USA}
\altaffiltext{4}{The Graduate University for Advanced Studies (SOKENDAI), 2-21-1 Osawa, Mitaka, Tokyo 181-8588 Japan}
\altaffiltext{5}{Joint ALMA Observatory, Alonso de Cordova 3107, Vitacura, Santiago 763-0355, Chile}
\altaffiltext{6}{Institute of Astronomy, University of Tokyo, 
2-21-1 Osawa, Mitaka, Tokyo 181-0015, Japan}
\altaffiltext{7}{Research Center for the Early Universe (WPI), University of Tokyo, 7-3-1 Hongo, Bunkyo, Tokyo 113-0033, Japan}
\altaffiltext{8}{Harvard-Smithsonian Center for Astrophysics, 60 Garden St., Cambridge, MA 02138, USA}
\altaffiltext{9}{Department of Astronomy, Graduate School of Science, The University of Tokyo, 7-3-1 Hongo, Bunkyo-ku, Tokyo 113-0033, Japan}
\altaffiltext{10}{National Astronomical Observatory of Japan, 2-21-1 Osawa, Mitaka, Tokyo 181-8588, Japan}

\begin{abstract}
We report deep ALMA observations complemented
with associated {\sl HST} imaging for a luminous ($m_{\rm UV}=25$) galaxy, 
`Himiko',  at a redshift z=6.595. The galaxy is remarkable for its 
high star formation rate, $100 M_\odot$yr$^{-1}$, securely
estimated from our deep {\sl HST} and {\sl Spitzer} photometry, and the
absence of any evidence for strong AGN activity or gravitational
lensing magnification.
Our ALMA observations probe an order of magnitude deeper
than previous IRAM observations, yet
fail to detect a 1.2mm dust continuum, indicating a flux $<52\mu$Jy 
comparable with or weaker than that of local dwarf irregulars with
much lower star formation rates. We likewise provide a strong upper limit 
for the flux of {\sc [Cii]} $158\mu$m, 
$L_{\rm [CII]} < 5.4\times 10^{7} L_\odot$,
a diagnostic of the hot interstellar gas often described as a valuable
probe for early galaxies. In fact, our observations indicate Himiko lies
off the local $L_{\rm [CII]}$ - star formation rate scaling relation by
a factor of more than 30. Both aspects of our ALMA observations
suggest Himiko is an unique object with a very low dust content
and perhaps 
nearly primordial
interstellar gas.  Our {\sl HST} images
provide unique insight into the morphology of this remarkable source, highlighting an 
extremely blue core of activity and two less extreme associated clumps.
Himiko is undergoing a triple major merger event whose 
extensive ionized nebula of Lyman alpha emitting gas, discovered
in our earlier work with Subaru, is powered by star formation and
the dense circum-galactic gas.
We are likely witnessing an early massive galaxy during a key
period of its mass assembly close to the end of the reionization 
era. 

\end{abstract}


\section{Introduction}
\label{sec:introduction}

Much progress has been achieved in recent years in charting the
abundance and integrated properties of the earliest galaxies beyond a redshift
of $z\simeq 6$ selected via optical and near-infrared (NIR)
photometry (e.g. \citealt{bouwens2010a,mclure2010,castellano2010,ouchi2010,ellis2013,mclure2012,schenker2012}).
The emerging picture indicates that the redshift period
$6\lesssim z\lesssim 10$ was a formative one 
in the assembly history of normal galaxies.
Sources at $z\simeq 7-8$ show moderately blue ultraviolet continua
possibly consistent with young, metal-poor stellar populations
with a star-formation rate (SFR) of $1-10 M_\odot$ yr$^{-1}$ 
(e.g. \citealt{bouwens2010b,finkelstein2010,schaerer2010,dunlop2012}). 
Their small physical sizes ($\simeq 0.7$ kpc; \citealt{oesch2010,ono2013}) and modest
stellar masses ($10^8-10^9 M_\odot$; \citealt{labbe2010})
suggest they quickly merge into larger, more luminous systems. 
The abundance
of sub-luminous, small galaxies at high redshift 
also indicates significant merging occurred at early times, given the
faint-end slope of the UV luminosity function changes from a steep $\alpha\simeq -1.9$ at 
$z=7-8$ \citep{schenker2012,mclure2012} to $\alpha\simeq -1.7$ at $z=2-3$ (e.g. \citealt{reddy2009}).

In practice it is hard to decipher the physical processes that
govern the early assembly of galaxies from 
integrated properties alone. We therefore seek to complement
statistical measures such as star formation rates and stellar
masses by detailed evidence from well-studied individual examples.
Likewise, our understanding of early cosmic history may
be incomplete given so much is currently deduced from optical and
near-infrared data alone\citep{robertson2013}. Although optical
and near-infrared selected sources at high redshift suggest they
contain little or no dust
\citep{bouwens2012,dunlop2013}, 
this may be a selection bias. Star formation 
obscured by dust cannot be quantified without identifying cold dust 
emission. Furthermore, the gas phase metallicity remains a key measurement for
understanding early systems, most notably in locating the highly-prized pristine
`first generation' systems unpolluted by supernova enrichment.
Neither optical nor near-infrared facilities can currently address
this important quest given the diagnostic metal lines used at lower redshift, such as 
{\sc [Oii]}$\lambda\lambda$3726,3729 \AA\ and {\sc [Oiii]}$\lambda\lambda$5007,4959 \AA\ ,
cannot be measured beyond $z\simeq$5 until the launch of the James
Web Space Telescope.

It is for this reason that state of the art sub-millimeter facilities such as the
 Atacama Large Millimeter Array (ALMA) offer enormous promise. First, they 
 can quantify the possible bias in our current "optical" view of early galaxy formation 
 by 
detecting the hidden cold dust in
high redshift galaxies. Secondly, the CO/{\sc [Cii]}
158$\mu$m features prominent in star forming regions in the local
Universe offer a valuable tracer of metallicity at early times. Thus far, neither cold dust 
continuum nor these low-ionization tracers of metallicity have been observed beyond
$z\sim 6$ (\citealt{vieira2013}, \citealt{capak2011}, \citealt{riechers2010}, \citealt{coppin2010}).
Although a few QSOs have been observed at sub-mm wavelengths to $z=6.4-7.1$ 
\citep{maiolino2005,iono2006,walter2009,venemans2012,willott2013,wang2013}, 
the presence of a powerful 
AGN undoubtedly complicates any understanding of the physical conditions in their host galaxies. 

Detecting these important diagnostic signals  of dust and metallicity from typical $z\simeq 7$ galaxies is 
clearly a major observational challenge. Only upper limits on {\sc [Cii]} and submm continuum fluxes
have been presented so far for the abundant population of Lyman break galaxies (LBGs) 
and Ly$\alpha$ emitters (LAEs) at $z\sim 7$. These limits have come from deep exposures
with the Submillimetre Common-User Bolometer Array (SCUBA;\citealt{holland1999}) 
facility on the James Clerk Maxwell Telescope and Plateau de Bure
interferometric observations  (e.g. \citealt{ouchi2009a,walter2012,kanekar2013}). 
Very recently, one
$z=6.34$ source has been studied in this way following a comprehensive search 
for red objects in the Herschel HerMES blank field survey at $50-500\mu$m
\citet{riechers2013}. This source, HFLS3, has a very strong far-infrared continuum emission 
and prominent molecular/low-ionization lines. Its star formation rate, inferred from its far-infrared 
luminosity, is extremely high, $2900 M_\odot$ yr$^{-1}$. Clearly we need to understand
the context of this remarkable object by observing other sources at a similar redshift.

The present work is concerned with undertaking
such a study for an extraordinarily luminous star-forming galaxy which will hopefully
complement the study of HFLS3 by \citet{riechers2013}. \citet{ouchi2009a} reported 
the discovery of the star-forming galaxy at $z=6.595$, 'Himiko'
\footnote{
See \citet{ouchi2009a} for the meaning of this name.
}
,
with a {\sl Spitzer}/IRAC counterpart.
This source was identified
from an extensive 1 deg$^2$ optical survey for $z=6.6$ galaxies
in the UKIDSS/UDS field conducted with the Subaru telescope. The redshift
was subsequently confirmed spectroscopically using Keck/DEIMOS.
The unique features of this remarkable source are evident in comparison to the 
total sample of 207 galaxies at $z=6.6$ found in the panoramic Subaru survey. 
Not only is Himiko by far the most luminous example
($M_{\rm UV}=25$; $L({\rm Ly}\alpha) = 4 \times 10^{43}$ erg s$^{-1}$),
but it is spatially extended in Ly$\alpha$ emission
whose largest isophotal area is 5.22 arcsec$^2$, corresponding to a linear
extent of over
17 kpc.
The lower limit, $SFR>34 M_\odot$yr$^{-1}$, is placed
on the SFR of Himiko by the spectral energy distribution (SED) fitting analysis
with the early photometric measurements
and the stellar-synthesis and nebular-emission models \citep{ouchi2009a}.
Due to the large uncertainties of photometric measurements,
\citet{ouchi2009a} cannot constrain $E(B-V)$, and
provide only the lower limit of SFR with $E(B-V)\ge 0$.

The present paper is concerned with the analysis of uniquely deep ALMA
and {\sl HST} observations of this remarkable source. Given its intense
luminosity and high star formation rate, the presumption
is that it is being observed at a special time in its assembly history.
We seek to use the cold dust continuum 
and {\sc [Cii]} measures from
ALMA to understand its dust content and gas phase metallicity.
Likewise the matched resolution of {\sl HST} will allow us to address its
morphologic nature. By good fortune,
one of the {\sl HST} intermediate band filters closely matches the
intense Lyman $\alpha$ emission observed for this source
with Subaru. Ultimately, we then seek to understand the physical
source of energy that powers the extensive Lyman $\alpha$
nebula.

A plan of the paper follows. We describe our ALMA and {\sl HST} observations
in \S \ref{sec:observations},
and present the detailed properties such as dust-continuum and metal-line emission, 
morphology, and stellar population in \S \ref{sec:results}.
We discuss the nature of this object
in \S \ref{sec:discussion}, and summarize our findings in \S \ref{sec:summary}.
Throughout this paper, magnitudes are in the AB system. 
We adopt 
$(h,\Omega_m,\Omega_\Lambda,n_s,\sigma_8)=(0.7,0.3,0.7,1.0,0.8)$.

\begin{deluxetable*}{cccccccc}
\tablecolumns{8}
\tabletypesize{\scriptsize}
\tablecaption{ALMA Observations and Sensitivities
\label{tab:alma_observations}}
\tablewidth{0pt}
\setlength{\tabcolsep}{0.02in}
\tablehead{
\colhead{$\nu_{\rm cont}$} &
\colhead{$\nu_{\rm line}$} &
\colhead{$\sigma_{\rm cont}$} &
\colhead{$\sigma_{\rm line}$} &
\colhead{$f_{\rm cont}$} &
\colhead{$f_{\rm line}$} &
\colhead{$L_{\rm FIR}$} &
\colhead{$L_{\rm [CII]}$} \\
\colhead{(GHz)} &
\colhead{(GHz)} &
\colhead{($\mu$Jy beam$^{-1}$)} &
\colhead{($\mu$Jy beam$^{-1}$)} &
\colhead{($\mu$Jy)} &
\colhead{($\mu$Jy)} &
\colhead{($10^{10} L_\odot$)} &
\colhead{($10^{7} L_\odot$)} \\
\colhead{(1)} &
\colhead{(2)} &
\colhead{(3)} &
\colhead{(4)} &
\colhead{(5)} &
\colhead{(6)} &
\colhead{(7)} &
\colhead{(8)} 
}
\startdata
259.007 & 250.239 & 17.4 & 83.3 & $<52.1$ & $<250.0$ & $<8.0$ & $<5.4$
\enddata
\tablecomments{
(1)-(2): Central frequencies of continuum and {\sc [Cii]} line observations that correspond to 
$1.16$ and $1.20$ mm, respectively.
(3)-(4): $1\sigma$ sensitivities for continuum and {\sc [Cii]} line in a unit of $\mu$Jy beam$^{-1}$. The continuum sensitivity is given in the total bandwidth for the continuum measurement is $19.417$ GHz or $86.894$ $\mu$m that is a sum of 4 spectral windows (see text). The line sensitivity is defined with a channel width of $200$ km$^{-1}$.
(5)-(6): $3\sigma$ upper limits of continuum and {\sc [Cii]} line in a unit of $\mu$Jy.
(7)-(8): $3\sigma$ upper limits of far-infrared continuum luminosity ($8-1000\mu$m) and {\sc [Cii]} line luminosity in a unit of $10^{10}$ and $10^{7}$ solar luminosities, respectively.
We estimate $3\sigma$ upper limits of far-infrared continuum luminosities at $40-500\mu$m and $42.5-122.5\mu$m
to be $<7.36\times 10^{10}$ and $<6.09\times 10^{10}L_\odot$, respectively. These far-infrared luminosities
are estimated with the assumptions of the graybody, $\beta_{\rm d}=1.5$, and the dust temperature of $T_d=40$K.
}
\end{deluxetable*}

\section{Observations and Measurements}
\label{sec:observations}

\subsection{ALMA}
\label{sec:alma_observations}

To understand whether obscured star-formation is an important issue as well
as the metallicity of Himiko, a key source at high redshift,
we carried out deep ALMA Band 6 observations
in 2012 July 15, 18, 28, and 31
with 16 12m-antenna array under the the extended configuration of 
36-400m baseline. 
The precipitable water vapor (PWV) ranged 
from 0.7 to 1.6 mm during the observations.
We targeted Himiko's {\sc [Cii]} line
of rest-frame 1900.54 GHz (157.74$\mu$m)
which is redshifted to 250.24 GHz (1.198mm) at a redshift of $z_{\rm Ly\alpha}=6.595$. 
Because a brighter dust continuum is expected
at a higher frequency in the 1.2mm regime, we extended
our upper sideband (USB) to the high-frequency side.
Thus, we targeted the {\sc [Cii]} line with the 
lowest spectral window (among 4 spectral windows)
in the lower sideband (LSB) and
set the central frequency of the 4 spectral windows
are 250.24 and 252.11 GHz in LSB, and
265.90 and 267.78 GHz in USB with a bandwidth of
1875 MHz. The two spectral windows and each sideband
cover the frequency ranges contiguously.
The total on-source integration time was 3.17 hours.
We used 3c454.3 and J0423-013 for bandpass calibrators and
J0217+017 for a phase calibrator.
The absolute flux scale was established by observations of Neptune and
Callisto.
Our data were reduced with Common Astronomy Software Applications (CASA) 
package. We rebin our data to a resolution of 166 MHz (200 km$^{-1}$).
The FWHM beam size of the final image is $0''.82 \times 0''.58$ with a position angle of $79^{\circ}.5$.
The $1\sigma$ noise of continuum image is $\sigma_{cont}=17.4$ $\mu$Jy beam$^{-1}$ 
over the the total bandwidth of $19.417$ GHz whose $7.5$ GHz is
sampled. The $1\sigma$ noise of {\sc [Cii]} line image
is 
$\sigma_{line}=83.3$ $\mu$Jy beam$^{-1}$ 
at 250.239 GHz over a channel width of $200$ km$^{-1}$.

Further details of the ALMA observations and sensitivities are summarized in
Table \ref{tab:alma_observations}.

\begin{figure}
\epsscale{1.0}
\plotone{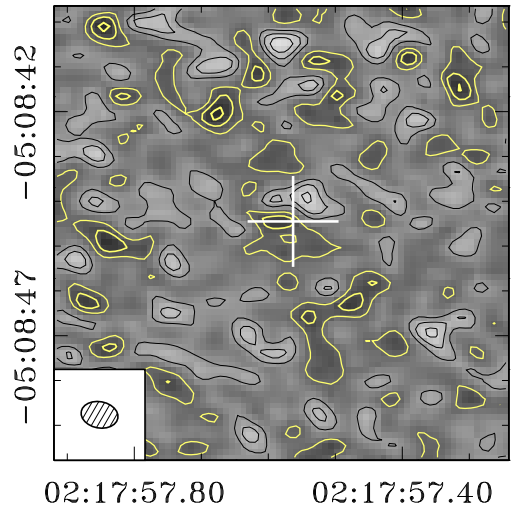}
\caption{
ALMA continuum data for Himiko at 259 GHz (1.16mm).
The gray scale indicates the intensity 
at each position where darker regions imply
higher intensities. The black contours denote
$-3$, $-2$, and $-1$ $\sigma$ levels, while yellow contours
show $+1$, $+2$, and $+3$ $\sigma$ significance levels,
where the $1 \sigma$ flux corresponds to $17.4\mu$Jy beam$^{-1}$.
The white cross indicates the position of Himiko. The ellipse in the
lower corner denotes the beam size.
\label{fig:himiko_cont1200um}}
\end{figure}

We averaged fluxes over the two spectral windows of LSB
(249.30-253.05 GHz or 1.203-1.185mm) and 
USB (264.96-268.71 GHz or 1.131-1.116mm) 
in the range of frequency free from the {\sc [Cii]} line.
Figure \ref{fig:himiko_cont1200um} presents the resulting
ALMA continuum data at $259.01$ GHz in frequency
(or 1.167mm in wavelength) with
a $1\sigma$ sensitivity of $17.4\mu$Jy beam$^{-1}$.
There is a $\sim 2\sigma$ flux peak in the beam size 
on the position of Himiko. However, there are 
a series of negative pixels nearby that correspond
to the $2-3 \sigma$ level per beam. We conclude therefore that
Himiko remains undetected in the 1.2mm continuum with a
$3\sigma$ upper limit is $<52.1\mu$Jy beam$^{-1}$.
We note that this sensitivity is two and one order(s) of magnitudes
better than those previously obtained by deep SCUBA/SHADES
and IRAM/PdBI observations \citep{ouchi2009a,walter2012}.
This clearly indicates Himiko has very weak millimeter emission.
Table \ref{tab:alma_observations} summarizes the flux upper limits for the
continuum and {\sc [Cii]} line derived
from our ALMA data.

\begin{figure}
\epsscale{1.15}
\plotone{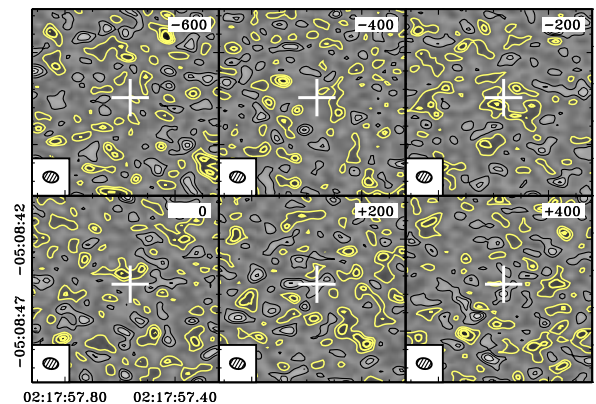}
\caption{
As Figure \ref{fig:himiko_cont1200um},
but for {\sc [Cii]} velocity channel maps of Himiko
whose $1\sigma$ intensity is $83.3 \mu$Jy beam$^{-1}$. 
The six panels present maps of 200 km$^{-1}$ width
at central velocities of $-600$, $-400$, $-200$,
$0$, $+200$, and $+400$ km s$^{-1}$ from the top left to
the bottom right. $0$ km s$^{-1}$
corresponds to {\sc [Cii]} emission at the redshift $z_{\rm Ly\alpha}=6.595$, 
i.e. 250.24 GHz (1.198mm).
\label{fig:himiko_lineCII}}
\end{figure}

Figure \ref{fig:himiko_lineCII} shows the ALMA {\sc [Cii]} velocity channel maps 
for Himiko. We searched for a signal of {\sc [Cii]} over
600 km $^{-1}$ around the frequency corresponding to $z_{\rm Ly\alpha}=6.595$.
In Figure \ref{fig:himiko_lineCII}, there are noise peaks barely reaching at the $3\sigma$ level
in 0 km s$^{-1}$ and $-200$ km s$^{-1}$ that are slightly north and south of 
the Himiko's optical center position,
respectively. However, neither of these sources is a reliable counterpart of Himiko.
Although the ALMA data reach a $1\sigma$ noise of $83.3\mu$Jy beam$^{-1}$,
no {\sc [Cii]} line is detected. 
The corresponding $3\sigma$ upper limit for {\sc [Cii]} 
is $250.0\mu$Jy in a channel width of 200 km s$^{-1}$. The associated luminosity
limit is $<5.4 \times 10^{7} L_\odot$.

\subsection{{\sl HST}}
\label{sec:hst_observations}

The primary goal of the associated {\sl HST} observations of
Himiko relate to the morphological nature of this remarkable
source. 
We carried out deep {\sl HST}/WFC3-IR
broad-band ($J_{125}$ and $H_{160}$)
\footnote{
$J_{125}$ and $H_{160}$
are referred to as $F125W$ and $F160W$, respectively.
and medium-band ($F098M$) 
observations for Himiko. 
The two broad-band filters of $J_{125}$ and $H_{160}$
measure the rest-frame UV continuum fluxes,
and are free from contamination from Ly$\alpha$ emission, 
}
thus maximizing the information content
on its stellar content for SED
fitting (\S \ref{sec:stellar_population}). The intermediate-band filter of 
$F098M$ fortuitously includes the spectroscopically-confirmed Ly$\alpha$ line of Himiko at 
9233\AA\ \citep{ouchi2009a} with a system throughput of 40\%, close to the peak 
throughput of this filter ($\sim 45$\%). 
Thus, the $F098M$ image is ideal for for mapping the distribution of Ly$\alpha$ emitting gas.

Our observations were conducted
in 2010 September 9, 12, 15-16, 18, and 26
with an ORIENT of 275 degrees.
Some observations were partially lost because {\sl HST} went into `safe mode'  on 2010 September 9, 
22:30 during the execution of one visit.
The total integration times for usable imaging data are 15670.5, 13245.5, 18064.6 seconds for
$F098M$, $J_{125}$, and $H_{160}$, respectively.
The various WFC3 images were reduced with
WFC3 and MULTIDRIZZLE packages on PyRAF.
To optimize our analyses, in the multidrizzle processing we 
chose a final$\_$pixfrac$=0.5$ and pixel scale of $0''.05132$.
We degraded images of $F098M$ and $J_{125}$ 
to match the PSFs of these images with 
the one of $H_{160} $that has the largest size among the {\sl HST} images.
We ensured the final WFC3 images have a matched PSF
size of $0''.19$ FWHM.

\begin{figure}
\epsscale{1.0}
\plotone{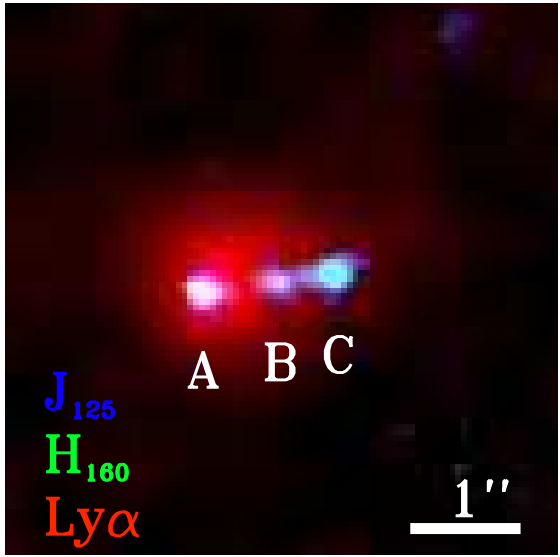}
\caption{
Color composite image of Himiko. Blue and green represent
{\sl HST}/WFC3 continua of $J_{125}$ and $H_{160}$, respectively.
Red indicates Ly$\alpha$ emission resolved with 
sub-arcsec seeing Subaru observations.
The Ly$\alpha$ emission image comprises the Subaru $NB921$ narrowband data with a subtraction
of the continuum estimated from the seeing-matched 
{\sl HST}/WFC3 data.
The three continuum clumps are labeled A, B and C.
\label{fig:hst_z7blob_composite_alma}}
\end{figure}

\begin{figure}
\epsscale{1.2}
\plotone{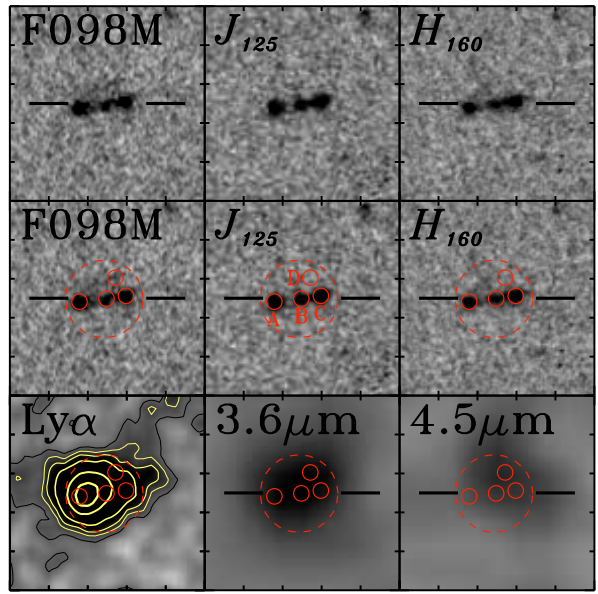}
\caption{
{\sl HST}, Subaru, and {\sl Spitzer} images of Himiko;
north is up and east is to the left.
Each panel presents $5''\times5''$ images at 
$F098M$, $J_{\rm 125}$, and $H_{\rm 160}$ bands
from {\sl HST}/WFC3,
$3.6\mu$m and $4.5\mu$m bands 
from {\sl Spitzer} SEDS. The Ly$\alpha$ image is a Subaru 
$NB921$ image continuum subtracted using $J_{\rm 125}$
and includes intensity contours. The Subaru image
has a PSF size of $0".8$.
The red-solid circles indicate
the positions of $0''.4$-diameter apertures 
for Clumps A, B, C, and D photometry in the {\sl HST} images
(see Section \ref{sec:hst_observations} for details), while the red-dashed circles denote 
$2''$-diameter apertures used for the defining the total magnitudes.
\label{fig:HimikoHstSpitzerSubaru_images}}
\end{figure}

Figure \ref{fig:hst_z7blob_composite_alma} presents
a color composite {\sl HST} UV-continuum image of Himiko,
together with a large ionized Ly$\alpha$ cloud
identified by the Subaru observations \citep{ouchi2009a}.
This image reveals that the system comprises
3 bright clumps of starlight surrounded 
by a vast Ly$\alpha$ nebula $\gtrsim 17$ kpc across. We
denote the three clumps as A, B, and C.
Figure \ref{fig:HimikoHstSpitzerSubaru_images} 
shows the {\sl HST}, Subaru, and {\sl Spitzer} images separately.
The $F098M$ image in Figure \ref{fig:HimikoHstSpitzerSubaru_images} 
detects only marginal extended Ly$\alpha$ emission, because of
the shallower surface brightness limit of the 2.4m {\sl HST} 
compared to the 8m Subaru telescope. Nevertheless,
we have found a possible bright extended component
at position D in Figure \ref{fig:HimikoHstSpitzerSubaru_images}.
We perform $0''.4$-diameter aperture photometry for the clumps A-C and 
location D as well as $2''$-diameter aperture photometry which we adopt
as the total magnitude of the system. 
Tables \ref{tab:properties_of_subcomponents} and
\ref{tab:total_magnitudes}
summarize the photometric properties. 
It should be noted that Himiko is not only identified as
an LAE, but also would be regarded as a LBG or `dropout' galaxy. 
Using the optical photometry of \citet{ouchi2009a}
(see also Table \ref{tab:total_magnitudes}),
we find no blue continuum fluxes for the filters from $B$ through $i'$ to the 
relevant detection limits of $28-29$ mag. The
very red color of $i'-z'>2.1$ meets typical dropout selection
criteria (e.g. \citealt{bouwens2011}). 
Because the $z'$-band photometry includes
the Ly$\alpha$ emission line and a Ly$\alpha$-continuum break, we can 
also estimate the continuum-break color using our {\sl HST} photometry
of $J_{125}$ and $H_{160}$ and the optical $i$-band photometry. Assuming
the continuum spectrum is flat ($f_\nu=$const.), we obtain a continuum break 
color $i'-J_{125}>3.0$ or $i'-H_{160}>3.0$, further supporting that Himiko as a 
LBG. Importantly, these classifications apply also to the clumps A-C ruling out
that some could be foreground sources. 

\begin{deluxetable*}{lcccccccccc}
\tablecolumns{11}
\tabletypesize{\scriptsize}
\tablecaption{Properties of the Subcomponents
\label{tab:properties_of_subcomponents}}
\setlength{\tabcolsep}{0.03in}
\tablewidth{0pt}
\tablehead{
\colhead{Component} & 
\colhead{$x$(pix)} &
\colhead{$y$(pix)} & 
\colhead{$NB$} &
\colhead{$J_{125}$} &
\colhead{$H_{160}$} &
\colhead{$\beta$} &
\colhead{$L({\rm Ly\alpha})$} &
\colhead{$EW_0$} &
\colhead{SFR(UV)} &
\colhead{SFR$({\rm Ly\alpha})$} \\
\colhead{} & 
\colhead{} &
\colhead{} & 
\colhead{(mag)} &
\colhead{(mag)} &
\colhead{(mag)} &
\colhead{} &
\colhead{($10^{42}$erg s$^{-1}$)} &
\colhead{(\AA)} &
\colhead{($M_\odot$yr$^{-1}$)} &
\colhead{($M_\odot$yr$^{-1}$)} \\
\colhead{(1)} & 
\colhead{(2)} &
\colhead{(3)} & 
\colhead{(4)} &
\colhead{(5)} &
\colhead{(6)} &
\colhead{(7)} &
\colhead{(8)} &
\colhead{(9)} &
\colhead{(10)} &
\colhead{(11)}
}
\startdata
Total$^{a}$ & 2815.4 & 2790.4 & $23.55\pm 0.05$ & $24.99\pm 0.08$ & $24.99\pm 0.10$ & $-2.00\pm 0.57$ & $38.4\pm 1.5$ & $78^{+8}_{-6}$ & $30\pm 2$ & $35\pm 1$ \\
\ \ \ \ \ \ \ (F098M) & \nodata & \nodata & ($24.84\pm 0.08$) & \nodata & \nodata & \nodata              & ($30.5\pm 15.6$) & ($61^{+28}_{-23}$) & \nodata & ($28\pm 14$) \\
A(clump)       & 2803.0 & 2789.0 & $26.36\pm 0.04$ & $26.54\pm 0.04$ & $26.73\pm 0.06$ & $-2.84\pm 0.32$ & $8.1\pm 1.9$  & $68^{+14}_{-13}$ & $7\pm 0$ & $7\pm 2$\\
B(clump)       & 2816.5 & 2790.5 & $27.19\pm 0.09$ & $27.03\pm 0.07$ & $27.04\pm 0.08$ & $-2.04\pm 0.47$ & $0.2\pm 1.9$  & $3^{20}_{-18}$ & $5\pm 0$ & $0\pm 2$\\
C(clump)       & 2826.3 & 2791.9 & $26.57\pm 0.05$ & $26.43\pm 0.04$ & $26.48\pm 0.05$ & $-2.22\pm 0.28$ & $0.8\pm 1.9$  & $6^{+12}_{-10}$ & $8\pm 0$ & $1\pm 2$\\
D             & 2821   & 2801   & $28.47\pm 0.29$ & $>28.86$        & $>28.70$        & \nodata         & $>1.7$        &  $>123$      & $<1$     & $>2$\\
\enddata
\tablecomments{
(1): Name of the component. (2)-(3): Positions in pixels. 
(4)-(6): Aperture magnitudes in $NB$, $J_{125}$, and $H_{160}$. $NB$ indicates
a narrow/intermediate band magnitude determined with $F098M$ 
for all lines except in Total(NB921).
The upper limit corresponds to a $3\sigma$ limit.
(7): UV-continuum slope.
(8): Ly$\alpha$ luminosity.
(9): Rest-frame apparent equivalent width of Ly$\alpha$ emission line in \AA. 
(10): Star-formation rate estimated from the UV magnitude.
(11): Star-formation rate estimated from the UV magnitude.
}
\tablenotetext{a}{
\vspace{0mm}
The $NB$ magnitude corresponds to $NB921$.
The quantities of (8)-(9) and (11) are estimated 
from $NB921$ photometry. 
}
\end{deluxetable*}

The UV continuum magnitudes of clumps A-C range
from 26.4 to 27.0 magnitudes in $J_{125}$ and $H_{160}$.
Each clump has a UV luminosity corresponding to the characteristic luminosity
$L^{\ast}$ of a $z\sim 7$ galaxy, $m$=$26.8$ mag
\citep{ouchi2009b,bouwens2011}). Moreover, the variation in
luminosity across the components is small; there is no single dominant point source in this system,
confirming earlier deductions that the system does not contain an active
nucleus. 

The $F098M$ image shows that Ly$\alpha$ emission is not
uniformly distributed across the 3 clumps.
Clump A shows intense Ly$\alpha$ emission 
with a rest-frame equivalent width ($EW_0$) of $68^{+14}_{-13}$\AA\
placing it in the category of a Lyman alpha Emitter (LAE), whereas clumps B and C are 
have emission more typical of Lyman break galaxies (LBGs) with a rest-frame Ly$\alpha$ 
equivalent width ($EW_0$) less than 20\AA\ given the measurement uncertainties.

In summary, the {\sl HST} and Subaru data indicates Himiko is a triple $L^*$ galaxy system 
comprising one LAE and two LBGs surrounded by an extensive 17 kpc diffuse Ly$\alpha$ halo. 
Importantly, from the above morphological studies, the possibility that Himiko is gravitational lensed
by a foreground concentration can be readily eliminated. Already, \citet{ouchi2009b} made
a strong case against lensing given the Keck spectroscopy revealed a velocity gradient of 
$60$km s$^{-1}$ across the system. We can further reject this supposition given there are
clear asymmetries in the outermost images (one has strong Ly$\alpha$ emission
and the other does not). 

\subsection{{\sl Spitzer}}
\label{sec:spitzer_data}

Although {\sl Spitzer} cannot match the resolution of the above morphological data, 
we use the very deep {\sl Spitzer}/IRAC SEDS data 
reaching 26 mag at the $3\sigma$ level
\citep{ashby2013}
to investigate the counterpart of the overall Himiko system at $3.6\mu$m
and $4.5\mu$m bands. 
To improve the relative astrometric accuracy,
we have re-aligned the SEDS images
to the {\sl HST} images, referring bright stellar objects
commonly detected in the Sptizer and {\sl HST} images.
The relative astrometric errors are estimated to be
$\simeq 0''.1$ rms.
We obtain total magnitudes of {\sl Spitzer}/IRAC images
from a $3''$-diameter aperture and use an aperture
correction given in \citet{yan2005}.
The total magnitudes are $23.69\pm 0.09$ mag
and $24.28\pm 0.19$ mag at $3.6\mu$m
and $4.5\mu$m bands, respectively.
Because the {\sl Spitzer}/IRAC $5.8\mu$m and $8.0\mu$m
and {\sl Spitzer}/MIPS $24\mu$m band images are not
available in the SEDS data set, we use the relatively
shallow {\sl Spitzer}/SpUDS (Dunlop et al.) photometry
measurements presented in \citet{ouchi2009a}.
Table \ref{tab:total_magnitudes} summarizes these 
total magnitudes and fluxes.

\begin{deluxetable}{cc}
\tabletypesize{\scriptsize}
\tablecaption{Total Magnitudes and Fluxes of Himiko
\label{tab:total_magnitudes}}
\tablewidth{0pt}
\setlength{}{}
\tablehead{
\colhead{Band} &
\colhead{Mag/Flux(Total)\tablenotemark{1}}\\
}
\startdata
$B$\tablenotemark{3} & $>28.7$ \\
$V$\tablenotemark{3} & $>28.2$ \\
$R$\tablenotemark{3} & $>28.1$ \\
$i'$\tablenotemark{3} & $>28.0$ \\
$z'$\tablenotemark{3} & $25.86\pm 0.20$ \\  
$NB921$\tablenotemark{3} & $23.55\pm0.05$ \\
$F098M$ & $24.84\pm 0.08$\\
$J_{125}$ & $24.99\pm 0.08$\\
$H_{160}$ & $24.99\pm 0.10$\\
$J$\tablenotemark{2} & $25.03\pm 0.25$ \\
$H$\tablenotemark{2} & $26.67\pm 2.21$ \\
$K$\tablenotemark{2} & $24.77\pm 0.29$ \\
$m(3.6\mu{\rm m})$ & $23.69\pm 0.09$ \\
$m(4.5\mu{\rm m})$ & $24.28\pm 0.19$ \\
$m(5.8\mu{\rm m})$\tablenotemark{3} & $>22.0$ \\
$m(8.0\mu{\rm m})$\tablenotemark{3} & $>21.8$ \\
$m(24\mu{\rm m})$\tablenotemark{3} & $>19.8$ \\
$S(1.2mm)$ & $<52.1\mu$Jy\tablenotemark{4}\\
$f({\rm [CII]})$ & $<250.0\mu$Jy\tablenotemark{4}\\
\enddata

\tablenotetext{1}{In units of AB magnitudes, if not specified.
The upper limits are $2\sigma$ and $3\sigma$ magnitudes in $BVRi'$ and
$5.8-24\mu$m bands, respectively.}
\tablenotetext{2}{Total magnitudes from UKIDSS/UDS DR8 data that are estimated with
a $2''$ diameter aperture photometry and the aperture
correction in the same manner as \citet{ono2010b}.
}
\tablenotetext{3}{Measurements obtained in \citet{ouchi2009a}.
The continuum magnitudes 
from $B$ through $z'$
are defined 
with a $2''$ diameter aperture photometry that
matches to the photometry of the total magnitudes
of NIR bands.
}
\tablenotetext{4}{Three sigma upper limits derived with
our ALMA data.}
\end{deluxetable}

\section{Results}
\label{sec:results}

\citet{ouchi2009a} found that Himiko has
a high SFR ($>34M_\odot$yr$^{-1}$)
and derived a moderately high stellar mass ($0.5-5.0\times 10^{10} M_\odot$) 
from the Subaru photometry and shallow {\sl Spitzer}/SpUDS data.
Here, we attempt to improve upon these estimates and, for the
first time, secure information on dust content and inter-stellar medium (ISM) metallicity.

\subsection{Far Infrared SED}
\label{sec:sed}

\begin{figure}
\epsscale{1.15}
\plotone{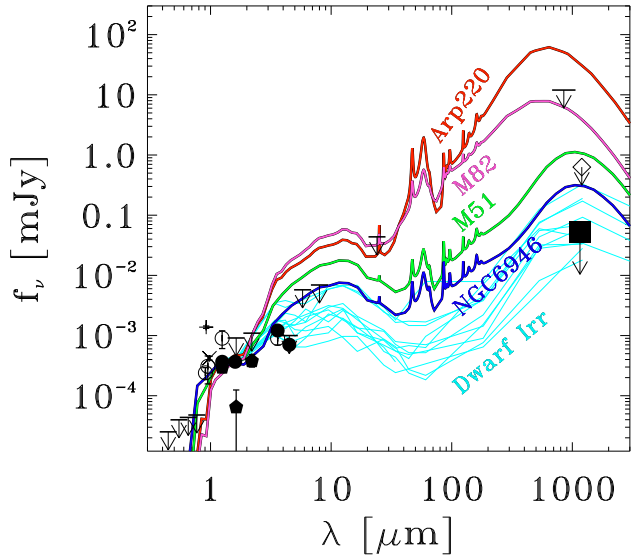}
\caption{
The optical to far-infrared SED of Himiko in the observed frame.
The filled square shows the upper limit from our deep ALMA Band 6 observations
and filled circles represent photometry from {\sl HST}/WFC3 $J_{125}$ and $H_{160}$
photometry and {\sl Spitzer} SEDS $3.6$ and $4.5\mu$m.
Filled pentagons indicate the UKIDSS-UDS DR8 $J$, $H$, and $K$ photometry.
Cross and plus symbols denote {\sl HST}/WFC3 $F098M$
and Suprime-Cam $NB921$ photometry that includes
Ly$\alpha$ emission and Gunn-Peterson trough
in their bandpasses. Open circles and arrows 
are data points and the upper limits  
taken from Ouchi et al. (2009). 
The open diamond with an arrow shows the upper limit
from the IRAM observations \citep{walter2012}.
Red, magenta, green, and blue lines represent
the SEDs of local galaxies, 
Arp220, M82, M51, and NGC6946 \citep{silva1998},
respectively, redshifted to $z=6.595$.
SEDs of local dwarf irregular galaxies similarly redshifted
are presented with cyan lines \citep{dale2007}.
All local galaxy SEDs are normalized 
in the rest-frame UV, where Himiko's SED
is determined reliably.
\label{fig:himiko_sed}}
\end{figure}

We investigate obscured star-formation and dust properties
of Himiko from its far-infrared (FIR) SED using the newly available
ALMA 1.2mm continuum data. The SED from the optical to millimeter
wavelengths is shown in Figure \ref{fig:himiko_sed}, together with 
that of various local starburst templates. The figure demonstrates that 
Himiko's millimeter flux is significantly weaker than that of
dusty starbursts in the local universe such as Arp220 and M82,
as well as the spiral galaxy NGC6946; it is more comparable to
those of dwarf galaxies of much lower mass. Similarly, Himiko's rest-frame
optical flux derived from the {\sl Spitzer}/IRAC $3.6$ and $4.5\mu$m photometry
is significantly weaker than those of dusty starbursts and spiral galaxies.
Given its intense rest-frame UV luminosity and moderately
high stellar mass, Himiko's dust emission and evolved stellar flux are remarkably 
weak. Both properties imply a low extinction and relatively young stellar age (\S3.3).
In this sense Himiko may be similar to many luminous $z\sim 3$ LBGs 
whose cold-dust continuum 
emission are also comparable to unreddened local starburst galaxies 
\citep{ouchi1999}.

We can estimate a far-infrared luminosity of Himiko
from our 1.2mm continuum limit. Assuming an optically thin graybody 
of modified blackbody radiation with a dust emissivity power-law spectral 
index of $\beta_{\rm d}=1.5$
and a dust temperature of $T_d=40$K \citep{eales1989,klaas1997}, 
we obtain a $3\sigma$ upper limit of $L_{\rm FIR}<8.0\times 10^{10} L_\odot$
integrated over $8-1000\mu$m. We also estimate $3\sigma$ upper limits of 
$<7.4\times 10^{10}$ and $<6.1\times 10^{10}L_\odot$
at $40-500\mu$m and $42.5-122.5\mu$m, respectively.
Note that these upper limits depend upon the assumed
dust temperature and $\beta_{\rm d}$. For $T_d=25$K
and $T_d=60$K, the $3\sigma$ upper limit luminosities
in $8-1000\mu$m are $<2.7\times 10^{10}$ and $<3.0\times 10^{11}L_\odot$, respectively.
Similarly, for $\beta_{\rm d}=0$ and $\beta_{\rm d}=2$, 
the $3\sigma$ upper limit luminosities
in $8-1000\mu$m are $<3.5\times 10^{10}$ and $<1.2\times 10^{11}L_\odot$, respectively.

The foregoing upper luminosity limits do depend
somewhat on dust temperature and spectral index.
Based on the Herschel measurements, \citet{lee2012} find
that the average dust temperature
is $\sim 30$ K under $\beta_{\rm d}=1.5$ for a relatively high redshift ($z\sim 4$) 
LBGs with a luminosity of $L\gtrsim 2L^*$ comparable to Himiko.
In the local universe, the median dust temperatures are 33 K, 30 K, and 36 K, for E/S0,
Sb-Sbc, and infrared bright galaxies, respectively. \citep{sauvage1994,young1989}.
Recent numerical simulations have claimed that LAEs may have a relatively high
dust temperature, due to the proximity of dust to star-forming regions. However, even
in this case  the maximum temperature reaches only $T_d\simeq 40$ K \citep{yajima2012a}.
On the other hand Himiko's dust must be heated to some lower limit by
the cosmic microwave background (CMB) whose blackbody temperature scales
as $T^{z=0}_{\rm CMB} (1+z)$, where $T^{z=0}_{\rm CMB}$
is the temperature of present-day CMB, $T^{z=0}_{\rm CMB}=2.73$ K.
Assuming local thermal equilibrium between ISM of Himiko and CMB at $z=6.595$ \citep{dacunha2013}, 
this yields a lower limit of $T_d=21$ K. Thus, it is appropriate to consider a range
of $T_d\simeq 20-40$ K with $\beta_{\rm d}\simeq 1.5$. Because
the larger assumed dust temperature $T_d=40$ K with $\beta_{\rm d}=1.5$
provides a weaker upper limit, we adopt a conservative $3\sigma$ upper limit of
$L_{\rm FIR}<8.0\times 10^{10} L_\odot$ ($8-1000\mu$m).
Tables \ref{tab:alma_observations} and \ref{tab:total_magnitudes} present 
the $3\sigma$ upper limit of luminosity.

\begin{figure}
\epsscale{1.1}
\plotone{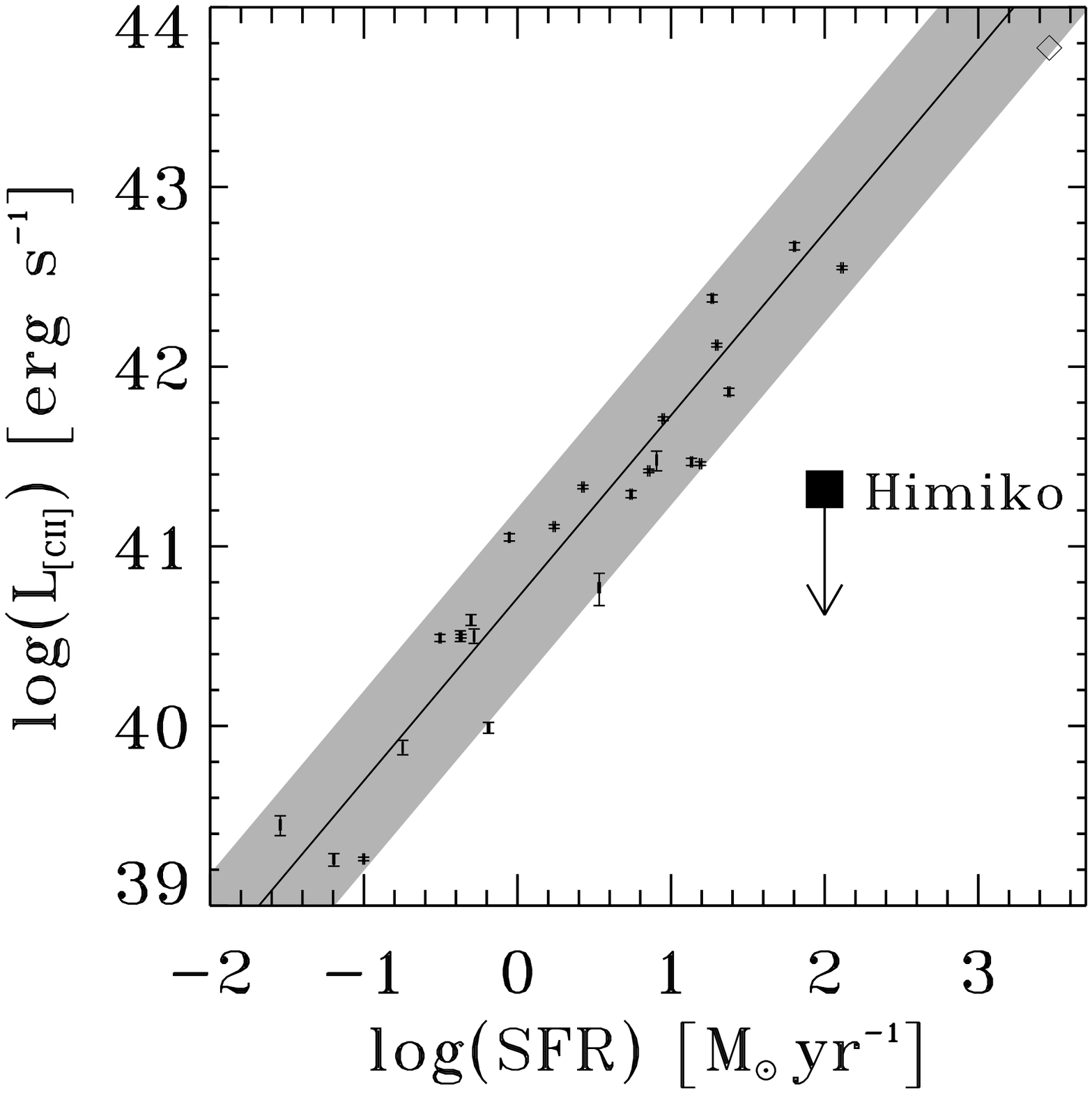}
\caption{
The {\sc [Cii]} luminosity as a function of SFR. The filled
square indicates the $3\sigma$ upper limit luminosity
of Himiko and the open diamond presents estimates of
HFLS3 \citep{riechers2013}. The solid line is the
local scaling relation determined with the data shown with bars
\citep{delooze2011}. Note that the bars are obtained
by re-calculation the SFR values using 
the data of \citet{delooze2011}, following the
formula shown in \citet{delooze2011}.
The shaded region indicates the observed scatter.
\label{fig:cii_sfr_himiko}}
\end{figure}

\begin{deluxetable*}{lcccccc}
\tablecolumns{7}
\tabletypesize{\scriptsize}
\tablecaption{Stellar Population of Himiko
\label{tab:stellar_population}}
\tablehead{
\colhead{Model} & 
\colhead{$M_{\rm *}$} &
\colhead{$E(B-V)_{*}$} & 
\colhead{Age} &
\colhead{$SFR$} &
\colhead{$sSFR$} &
\colhead{$\chi^2/$dof} \\
\colhead{} & 
\colhead{($M_\odot$)} &
\colhead{(mag)} & 
\colhead{Myr} &
\colhead{$M_\odot$yr$^{-1}$} &
\colhead{yr$^{-1}$} &
\colhead{} \\
\colhead{(1)} & 
\colhead{(2)} &
\colhead{(3)} & 
\colhead{(4)} &
\colhead{(5)} &
\colhead{(6)} &
\colhead{(7)} 
}
\startdata
stellar$+$nebular & $1.5^{+0.2}_{-0.2} \times 10^{10}$ & $0.15$\tablenotemark{a} & $182^{+22}_{-20}$ & $100\pm 2$ & 
$6.7 \pm 0.9 \times 10^{-9}$ & $1.55$ \\
pure stellar & $3.0^{+0.4}_{-0.6} \times 10^{10}$ & $0.15$\tablenotemark{a} & $363^{+44}_{-75}$ & $98\pm 2$ & 
$3.3 \pm 0.5 \times 10^{-9}$ & $3.13$
\enddata
\tablecomments{
(1): Models with or without nebular emission. 
(2): Stellar mass.
(3): Color excess of dust extinction for stellar continua.
(4): Stellar age.
(5): Star-formation rate.
(6): Specific star-formation rate.
(7): Reduced $\chi^2$. The degree of freedom (dof) is six.
}
\tablenotetext{a}{
The uncertainty of color excess is smaller than our model-parameter grid of 
$\Delta E(B-V)=0.01$.
}
\end{deluxetable*}

\subsection{ISM metallicity from {\sc [Cii]} Emission}
\label{sec:cii_emission}

We now turn to estimating the metallicity of the ISM of Himiko using with {\sc [Cii]} emission 
as a valuable tracer in star-forming regions. Despite our significant integration, no line is seen.
Figure \ref{fig:cii_sfr_himiko} (and Table \ref{tab:total_magnitudes}) 
presents the upper limit to the {\sc [Cii]} luminosity 
in the context of 
the correlation
with the star formation rate (SFR) \citep{delooze2011}. In the case of Himiko, the
SFR was obtained by SED fitting of the rest-frame UV to optical data including a correction for dust extinction 
(Section \ref{sec:stellar_population}). Himiko clearly departs significantly from the scaling relation; the deficit
amounts to a factor $\simeq\times$30. Given the SFRs of \citet{delooze2011}
for local galaxies 
are derived in a similar manner to that for Himiko, including contributions 
from dust-free and dusty starbursts with GALEX's UV and Spitzer's infrared fluxes, respectively, 
it seems difficult to 
believe this deficiency arises from some form of bias arising from comparing different populations.

\citet{gracia-carpio2011} and \citet{diazsantos2013} present $L_{\rm [CII]}/L_{\rm FIR}$ ratios 
for local starbursts that depend on $L_{\rm FIR}$ and the FIR and mid-IR surface brightnesses.
As a result, \citet{diazsantos2013} argue  that $L_{\rm [CII]}$ may not represent a particularly
reliable indicator of SFR. However, FIR and mid-IR luminosities only trace dusty starbursts and
typically exclude dust-free measures such as the UV luminosity. Because galaxies with 
fainter FIR/mid-IR luminosities have a larger ratio of $L_{\rm [CII]}/L_{\rm FIR}$ in the datasets
probed by  \citet{gracia-carpio2011} and \citet{diazsantos2013}, more dust-free star-formation
is expected in such systems. In this sense, the analysis of \citet{delooze2011} is perhaps more relevant
as a prediction of what to expect for Himiko.  Nonetheless, given the importance of using $L_{\rm [CII]}$ 
as a possible tracer and the discussion that follows below, independent studies of $L_{\rm [CII]}$
as a function of UV luminosity and $L_{\rm FIR}$ would be desirable.
Figure \ref{fig:cii_sfr_himiko} also shows that HFLS3 at $z=6.3$ \citep{riechers2013} follows the local 
scaling relation. However, it should be noted that the SFR of HFLS3 is derived from the far-infrared luminosity 
and thus any contribution from dust-free star-formation would be missing. In this sense, the SFR is possibly
a lower limit, in which case HFLS3 may also depart somewhat from the local relation.

The absence of {\sc [Cii]} emission in Himiko is perhaps the most surprising result from our ALMA
campaign. The emission line is often assumed to be the most robust far-IR tracer of star formation
in high redshift galaxies, such that it may replace optical lines such as Ly$\alpha$ in securing
spectroscopic redshifts in the reionization era. Our failure to detect this line in one of the most
spectacular $z\simeq$7 galaxies has significant implications which we discuss in Section 4.

\subsection{Improved Physical Properties from the Near-Infrared SED}
\label{sec:stellar_population}

Although some constraints on the integrated properties of Himiko were derived
in our earlier work \citep{ouchi2009a}, no $E(B-V)$ estimate
and only the lower limit of SFR with $E(B-V)\ge 0$ were obtained,
due to the large uncertainties of photometric measurements.
We now refine these estimates based
on our significantly deeper {\sl HST} and {\sl Spitzer} data. Our near-IR SED is taken using
total magnitudes from the {\sl HST} images (\S \ref{sec:hst_observations}), the 
{\sl Spitzer}/IRAC SEDS images (\S \ref{sec:spitzer_data}) and $JHK$ DR8 data from the
UKIDSS/UDS survey. We tabulate these total magnitudes in Table \ref{tab:total_magnitudes}
including ground-based optical data previously given in \citet{ouchi2009a}.

\begin{figure}
\epsscale{1.15}
\plotone{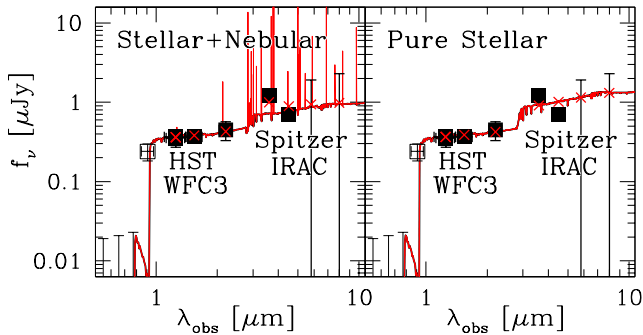}
\caption{
The optical to near-infrared SED of Himiko newly obtained
by our deep {\sl HST} and {\sl Spitzer} observations, together with
photometry from ground-based observations.
The red lines represent the best-fit SEDs of stellar synthesis models 
with (left) and without (right) nebular lines (see Ono et al. 2010 
for detailed model descriptions). 
The filled squares denote HST and Spitzer/SEDS 
fluxes of Himiko defined by the total magnitudes.
The open squares show $z'$-band fluxes that
are not used for the SED fitting, due 
to the Ly$\alpha$ line contamination.
The large error bars at $\le 0.8\mu$m and
$>5\mu$m are those obtained by 
Subaru and Spitzer/SpUDS observations
given by \citet{ouchi2009a}.
The red crosses represent the 
broadband fluxes expected from the best-fit
SED models.
For various assumptions the fits
indicate that Himiko has a SFR of $100M_\odot$ yr$^{-1}$, 
stellar mass of $2-3\times 10^{10} M_\odot$,
and a selective extinction of $E(B-V)=0.15$ (see text for details).
\label{fig:Himiko1}}
\end{figure}

We present the SED of Himiko in Figure \ref{fig:Himiko1} and undertake
$\chi^2$ fitting of a range of stellar synthesis models in the same manner 
as \citet{ono2010b} using the stellar synthesis models of \citet{bruzual2003} with dust attenuation
formulate given by  \citet{calzetti2000}. 
We adopt Salpeter initial mass function (IMF; \citealt{salpeter1955})
with lower and upper mass cutoffs of 0.1 and 100 $M_{\odot}$, respectively.
Applying models of constant and exponentially-decaying 
star-formation histories with
metallicities ranging from $Z=0.02-1.0Z_\odot$,
we search for the best-fit model in a parameter space of
$E(B-V)=0-1$ and age$=1-810$ Myr (where the latter upper limit corresponds to the cosmic age at $z=6.595$).
Nebular continuum and line emission, estimated from the ionizing photons from young stars, are optionally included
following the metallicity-dependent prescriptions presented in \citet{schaerer2009,ono2010b}. 

For a constant star-formation rate history with no nebular emission and
a fixed metallicity of $Z=0.2Z_\odot$, we find our best-fit model has 
a stellar mass of $M_*=3.0^{+0.4}_{-0.6}\times 10^{10}M_\odot$,
a stellar age of $3.6^{+0.4}_{-0.8}\times 10^{8}$yr, 
a SFR of $98^{+2}_{-2}M_\odot$yr$^{-1}$, and
extinction of $E(B-V)=0.15$
with a reduced $\chi^2$ of $3.1$.
This is a significant improvement over our much weaker earlier constraints which did not have
the benefit of the {\sl HST}/WFC3 or {\sl Spitzer}/SEDS data\citep{ouchi2009a}.
The new infrared data provide a critical role in determining the Balmer break 
thereby resolving the degeneracy between extinction and age.
On the other hand, the fit itself is not very satisfactory. The reduced $\chi^2$ is 
large and there is a significant discrepancy at $3.6\mu$m. Since the
$3.6\mu$m and $4.5\mu$m bands sample the strong nebular lines of 
H$\beta$+{\sc[Oiii]} and H$\alpha$, respectively, at $z=6.595$, this encourages
us to include nebular emission in our fitting procedure. In fact, in Figure 
\ref{fig:HimikoHstSpitzerSubaru_images}, we note the IRAC $4.5\mu$m 
emission shows a positional offset with respect to that at $3.6\mu$m suggesting
the possibility of contamination by nebular emission.

Adding nebular emission
to the stellar SED models given above,
the best fit has a more satisfactory reduced $\chi^2$, $1.6$,
and we derive a reduced stellar mass of $M_*=1.5^{+0.2}_{-0.2}\times 10^{10}M_\odot$,
a younger stellar age of $1.8^{+0.2}_{-0.2}\times 10^{8}$yr, but similar values for the 
SFR of $100^{+2}_{-2}M_\odot$yr$^{-1}$ and extinction of $E(B-V)=0.15$.
Table \ref{tab:stellar_population} summarizes 
the results of our SED fitting with
the pure stellar and stellar$+$nebular models.
In the stellar$+$nebular models, we assume that
all ionizing photons lead to nebular emission lines
corresponding to an escape fraction $f_{\rm esc}$=0.
If we allow $f_{\rm esc}$ to be a free parameter,
following \citet{ono2012} we find no change from
the model above (i.e. $f_{esc}$=0) and formally
establish that  $f_{\rm esc}<0.2$.

\citet{labbe2010} and \citet{finkelstein2010} have suggested
from their pure stellar models that {\sl HST} $z=7-8$ dropout galaxies
have modest stellar masses  ($10^8-10^9 M_\odot$) and are quite young ($30-300$Myr),
in contrast with Himiko's stellar mass ($M_*\simeq 3.0\times 10^{10}M_\odot$)
and age ($360$Myr) estimated with our pure stellar models.
Of course, Himiko is more massive and energetic than typical LBGs
seen in the small area of Hubble Ultra Deep Field. Its most notable feature
is its high SFR of $\simeq 100 M_\odot$yr$^{-1}$
which is more than an order of magnitude larger than those of the {\sl HST} LBGs
at similar redshifts ($\simeq 1-10 M_\odot$yr$^{-1}$; \citealt{labbe2010}).
Himiko's selective extinction, $E(B-V)=0.15$, is also larger than
that of {\sl HST} dropouts, more than half of which are consistent with
no extinction \citep{finkelstein2010}. On the other hand, the stellar mass of Himiko is 
only about 1/10th that of many submm galaxies (SMGs) at $z\sim 3$ \citep{chapman2005}.
We estimate a specific star-formation rate, $sSFR$, defined by a ratio
of star-formation rate to stellar mass to be
$sSFR=3.3 \pm 0.5 \times 10^{-9}$ and
$sSFR=6.7 \pm 0.9 \times 10^{-9}$ yr$^{-1}$,
for the pure stellar and stellar$+$nebular cases,
respectively.
Even though the stellar masses are very different, Himiko, SMGs and LBGs at $z\sim 3$ share comparable $sSFR$s 
 $\sim 10^{-9}-10^{-8}$ yr$^{-1}$ (see Figure 12 of \citealt{ono2010a}).

\begin{figure}
\epsscale{1.2}
\plotone{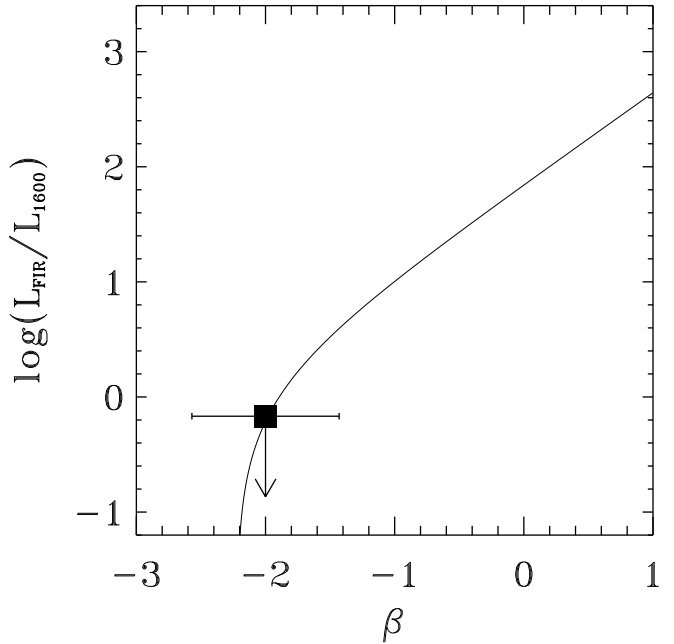}
\caption{
The UV to FIR luminosity ratio, $\log(L_{\rm FIR}/L_{\rm 1600})$,
as a function of the UV-continuum slope, $\beta$.
The filled square presents the upper limit of
$\log(L_{\rm FIR}/L_{\rm 1600})$ and the measurement of $\beta$ for 
the total luminosities of Himiko.
Solid line denotes the relation for local starburts
given by \citet{meurer1999}.
\label{fig:IRXbeta}}
\end{figure}

\subsection{UV Spectral Slopes on the Spatially Resolved Images}
\label{sec:uv_slopes}

The new {\sl HST} data gives us the first reliable measurement of the UV
continuum slope for each of the morphological components identified
in Figure \ref{fig:HimikoHstSpitzerSubaru_images}. 
The UV spectral slope provides a valuable indicator of 
the combination of dust extinction, metallicity, the upper IMF 
and stellar age. We estimated the UV slope, $\beta$, from the $J_{125}$ and $H_{160}$ 
photometry that samples the continua at the rest-frame wavelengths of
$\sim 1600$ and $\sim 2100$\AA\ neither of which is contaminated by either 
Ly$\alpha$ emission nor the Ly$\alpha$-continuum break.

We calculate $\beta$ via
\begin{equation}
\beta
       =-\frac{J_{125} - H_{160}}{2.5 \log \left( \lambda_{\rm c}^1 / \lambda_{\rm c}^2 \right)} - 2.
\label{eq:beta}
\end{equation}
where $\lambda_{\rm c}^1$ and $\lambda_{\rm c}^2$
are the central wavelengths of the $J_{125}$ and $H_{160}$ filters, respectively.
The estimates for each component are summarized in Table \ref{tab:properties_of_subcomponents}.
We obtain $\beta=-2.00\pm 0.57$ for the entire system of Himiko,
comparable to the average UV slope of $\simeq L^*$ LBGs, $\beta=-2.09\pm0.22$
(\citealt{bouwens2012}, see also \citealt{dunlop2013}).
Figure \ref{fig:IRXbeta} shows the UV to FIR luminosity ratio, 
$\log(L_{\rm FIR}/L_{\rm 1600})$, and the UV-continuum slope, $\beta$,
for the entire system of Himiko, and compares these estimates
with the relation of local starbursts \citep{meurer1999}.
Figure \ref{fig:IRXbeta} indicates that Himiko has
$\log(L_{\rm FIR}/L_{\rm 1600})$-$\beta$
values comparable with or smaller than those of local dust-poor starbursts.
Since the Small Magellanic Cloud (SMC) extinction 
has a smaller $\log(L_{\rm FIR}/L_{\rm 1600})$ value
at a given $\beta$ (see Figure 10 of \citealt{reddy2010})
due to SMC's steeper extinction curve in $A_{\lambda}/A_{\rm V}$-$1/\lambda$
than that for local starbursts, it may be more appropriate for Himiko.
Our result also suggests that Himiko is not associated with
additional FIR sources which are invisible in the rest-frame UV.
These implications are consistent with the conclusions of UV-FIR luminosity ratio 
discussed in Figure \ref{fig:himiko_sed}. 

\begin{figure}
\epsscale{1.2}
\plotone{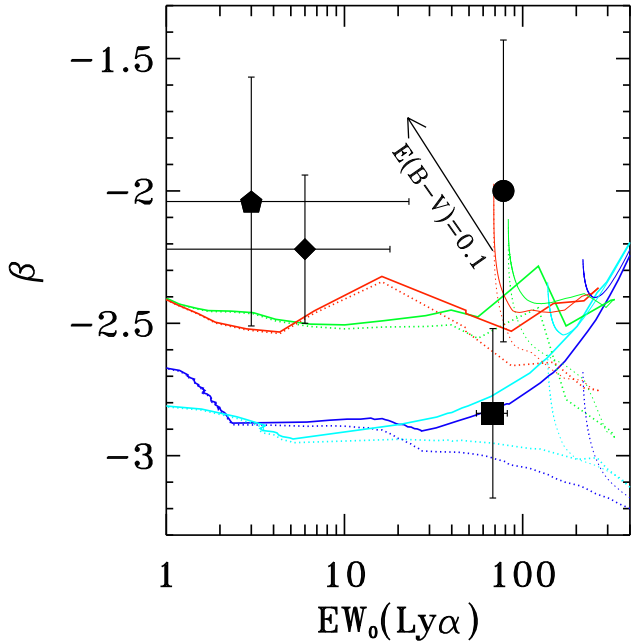}
\caption{
The UV-continuum slope, $\beta$, as a function of
rest-frame Ly$\alpha$ equivalent width.
The filled circle, square, pentagon, and diamond
denote measures for the entire system, Clumps A, B, and C,
respectively. Blue, cyan, green, and red solid lines
represent predictions based on instantaneous starburst models of 
PopIII, $Z=10^{-5}Z\odot$, $0.01Z\odot$, 
and $0.2Z\odot$ with nebular and stellar continua
\citep{raiter2010}.
Thin lines are the same, but
for constant star-formation models.
The associated dotted lines show the effect of ignoring
nebular emission. The arrow indicates the effect of applying 
an extinction with $E(B-V)_s=0.1$
(see the text for details).
\label{fig:BetaLya}}
\end{figure}

More interestingly, the UV slopes of the individual substructures provide valuable
information on the nature of Himiko. Clumps B and C have
$\beta=-2.04\pm 0.47$ and $\beta=-2.22\pm 0.28$, respectively,
comparable to the the average UV slope of $\simeq L^*$ LBGs.
However, Clump A presents a very blue UV slope, $\beta=-2.84\pm 0.32$.
Because this component is detected at the $\sim 20 \sigma$ level in both 
$J_{125}$ and $H_{160}$, the UV slope is quite reliable. \citet{bouwens2012} claim 
that selection and photometric biases lead to an error of only $\Delta \beta\simeq +0.1$ for
the brightest of their sources with $\sim 20\sigma$ photometry (see also \citealt{dunlop2013}).
Even including such a possible bias, Clump A remains significantly bluer than 
the average $\simeq L^*$ LBGs at the $\simeq 2\sigma$ level.

As presented in Section \ref{sec:hst_observations}, Clump A also 
shows Ly$\alpha$ emission. Together with the blue UV slope, this suggests 
a very young and/or metal poor component. However, the Ly$\alpha$ equivalent width is 
only $EW_0=68^{+14}_{-13}$\AA . To understand the significance of this, in
Figure \ref{fig:BetaLya}, we compare $\beta$ and $EW_0$ for the entire Himiko system 
and the various clumps with the stellar and nebular models of \citet{raiter2010},
where Salpeter IMF is assumed.
In Figure \ref{fig:BetaLya},
the arrow size in $\beta$ for the stellar extinction of $E(B-V)_s=0.1$
is calculated with the combination
of the empirical relation, $A_{1600}=4.43+1.99\beta$ \citep{meurer1999}, and 
Calzetti extinction, $A_{1600}=k_{1600} E(B-V)_s$, where $k_{1600}$ is $10$ \citep{ouchi2004a}.
Similarly, the arrow size in $EW_0$ for $E(B-V)_s=0.1$
is estimated from the relation 
given in \citet{ono2010a} under the assumption of $f_\nu$ flat continuum
and the standard starformation rate relations of UV and Ly$\alpha$ luminosities 
in the case B recombination.
Figure \ref{fig:BetaLya} shows that the data points of Himiko
fall on the tracks of star-formation photoionization models \citep{raiter2010} 
within the measurement errors and the dust-extinction correction uncertainties,
and indicates that Ly$\alpha$ emission of Himiko can be explained by
the photoionization by massive stars.

\section{Discussion}
\label{sec:discussion}

We now bring together our key results, both from the earlier Subaru program\citep{ouchi2009a}
and the present {\sl HST} and ALMA campaigns, in order to understand the significance
of our upper limits on the [C II] and dust emission and thereby the nature of Himiko.

\subsection{The Low Dust and Metal Content of Himiko}
\label{sec:weak_submm}

We have shown (Figure \ref{fig:himiko_sed}) that 
Himiko's submm emission is comparable with or weaker than 
that of local dwarf irregulars with far lower star-formation rates, 
indicating intensive star-formation in a dust-poor gaseous environment.
In fact, assuming the local starburst $SFR-L({\rm FIR})$ relation of \citet{kennicutt1998}
with the Himiko's FIR upper limit luminosity of $<8\times 10^{10}L_\odot$,
we obtain $SFR({\rm FIR})<14 M_\odot$yr$^{-1}$ that is 
far smaller than not only our best optical-NIR estimate SFR of 
$\simeq 100 M_\odot$yr$^{-1}$, but also
the UV-luminosity SFR of $SFR({\rm UV})=30\pm 2 M_\odot$yr$^{-1}$ 
with no dust extinction correction.
This is also true under the assumption of the $SFR-L({\rm FIR})$ relation
\citep{buat1996} valid for local dust poorer disk systems of Sb and later galaxies,
which provide $SFR({\rm FIR})<25 M_\odot$yr$^{-1}$.
In this way, Himiko does not follow the $SFR-L({\rm FIR})$ relation of typical
local galaxies, indicating a dust-poor gaseous environment.
This seems similar to observations which find extended Ly$\alpha$ emission 
in dust poor low-$z$ galaxies \citep{hayes2013} and a high-$z$
QSO \citep{willott2013}. 
Based on numerical simulations, \citet{dayal2010} find
that $z\sim 6-7$ LAEs are dust poor with a dust-to-gas mass
ratio smaller than Milky Way by a factor of 20.
\citet{dayal2010} predict a 1.4mm continuum flux of $\simeq 50\mu$Jy
for sources with $L({\rm Ly\alpha})=2-3\times 10^{43}$ erg s$^{-1}$ 
at $z=6.6$, a result comparable with our ALMA observations. 
Deeper ALMA observations could further test the model of
\citet{dayal2010} and place important constraints on
the dust-to-gas mass ratio.

\begin{figure}
\epsscale{1.2}
\plotone{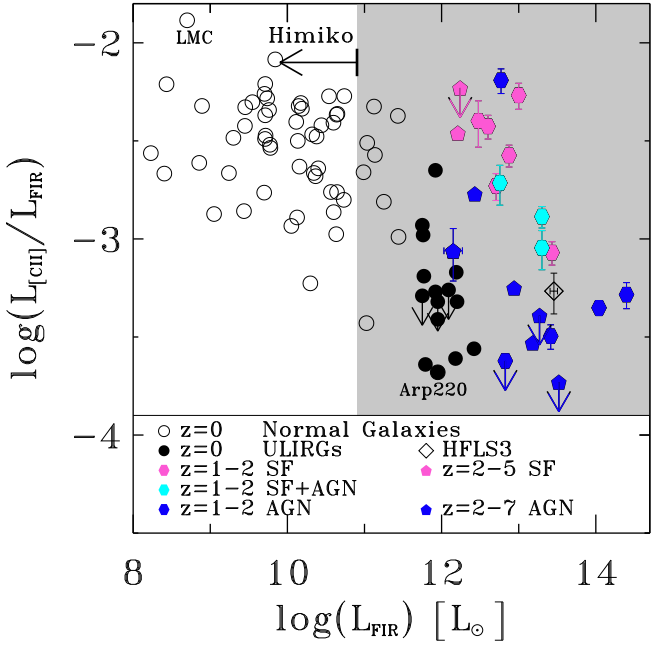}
\caption{
The ratio of {\sc [Cii]} luminosity to FIR luminosity 
as a function of FIR luminosity. The
thick bar with the arrow presents the
FIR upper limit for Himiko
at the arbitrary position in the vertical axis,
and the gray shaded region denotes the excluded range.
The ratios for star-forming (SF) and AGN-dominated
galaxies at $z=1-2$ are shown with magenta and
blue pentagons, respectively, while those of the
intermediate SF+AGN population at $z=1-2$
are represented with cyan hexagons \citep{stacey2010}.
Similarly, SF and AGN dominated galaxies at $z>2$ are
indicated by magenta pentagons and blue hexagons,
respectively \citep{marsden2005,maiolino2005,maiolino2009,
iono2006,pety2004,ivison2010,venemans2012}.
The black open diamond represents HFLS3 at $z=6.3$
\citep{riechers2013}.
Black open and filled circles denote local normal galaxies \citep{malhotra2001} 
and ULIRGs (\citealt{maiolino2005} and references therein), 
respectively, together with well-known local galaxies of 
the LMC, M82, and Arp220. 
\label{fig:cii_fir_himiko}}
\end{figure}

Similarly, our strong upper limit on the {\sc [Cii]} $158\mu$m line (Figure \ref{fig:cii_sfr_himiko}) 
places it significantly below the 
scaling relation of
$L_{\rm [CII]}$ and SFR obeyed by lower redshift galaxies.
This discovery indicates the following four possibilities:
Himiko has 
a) a hard ionizing spectrum from an AGN,
b) a very high density of photo-dissociation regions (PDRs),
c) a low metallicity, and
d) a large column density of dust.
In the case of a), a hard ionizing spectrum of AGN can produce
little {\sc [Cii]} luminosity relative to FIR luminosity,
due to the intense ionization field \citep{stacey2010}.
As we discuss in (ii) of Section 4.2, there are no signatures
of AGN; no detections of X-ray and high-ionization lines as well
as extended sources plus non-AGN like Ly$\alpha$ profile+surface brightness.
We can rule out the possibility of a).
In the case of b), a very high density of PDRs gives
more rapid collisional de-excitations for the forbidden line of {\sc [Cii]},
and quench a {\sc [Cii]} emission line. 
In the case of c), the PDRs in Himiko
are composed of metal poor gas that may be quite typical of
normal galaxies observed at early epochs. \citet{delooze2012} has argued that 
offsets from the {\sc [Cii]}-SFR relation can be 
explained in
terms of metal abundance and this would imply a gas-phase metallicity 
of $\lesssim 0.03Z_\odot$. Indeed, for our young mean stellar age of $160-410$ Myr, 
standard ionization-photon bounded {\sc Hii} regions with a local chemical abundance 
would yield {\sc [Cii]} emission somewhat above the local scaling relation,
due to the expected large PDRs.
Moreover, the recent numerical simulations predict that
a {\sc [Cii]} flux drops as metallicity decreases
\citep{vallini2013}. \citet{vallini2013} claim that
Himiko's gas-phase metallicity is sub-solar
with their models and the previous IRAM  {\sc [Cii]} upper limit. 
Comparing these numerical models 
with our strong ALMA upper limit of {\sc [Cii]}
would place further constraints on metallicity of Himiko.
In the case of d), the depth of C$^+$ zones in PDRs is determined
by dust extinction. Since the C$^+$ zones 
extend over the dust extinction up to $A_v\le 4$ \citep{malhotra2001}, 
heavy dust extinction in ISM does not 
allow to make a large C$^+$ zones emitting {\sc [Cii]}.
However, from no detection of 1.2 mm dust continuum discussed above,
the heavy dust extinction narrowing the PDRs is unlikely.
As dust extinction and gas phase metallicity generally correlate
closely (\citealt{storchi-bergmann1994}; see also \citealt{finlator2006}),
the weak dust emission also suggests a very low metallicity gas.
Thus, the case of c) is probably true, which contribute
the weak {\sc [Cii]} emission. The case of b) could also help
weakening the {\sc [Cii]} emission. To summarize, 
faint {\sc [Cii]} and weak dust emission can be explained
in a self-consistent manner 
with a very low metallicity gas and little dust in a near-primordial
system. 

It is informative to compare the above conclusion with the only other
well-studied galaxy at this redshift, HFLS3 at $z$=6.34 \citep{riechers2013},
recognizing that both it and Himiko were selected on the basis of
their extreme properties. Figure \ref{fig:cii_fir_himiko} presents the 
ratio of {\sc [Cii]} to the FIR luminosity as a function of FIR luminosity. Although 
Himiko is significantly offset from the trend shown by AGN and local starbursts,
this is not the case for HFLS3. Although \citet{riechers2013} claim that HFLS3 is 
free from AGN activities on the basis of the level of excitation for CO and H$_2$O,
its small {\sc [Cii]} to $L_{\rm FIR}$ ratio of $L_{\rm [CII]}/L_{\rm FIR}=5\times 10^{-4}$
suggests otherwise \citep{stacey2010,sargsyan2012}. 
On the other hand, the recent study of \citet{diazsantos2013} finds that
a luminous infrared galaxy (LIRG) with compact star-forming regions show a smaller $L_{\rm [CII]}/L_{\rm FIR}$ value,
and that a $L_{\rm [CII]}/L_{\rm FIR}$ value of pure star-forming LIRG drops by an order of
magnitude, from $10^{-2}$ to $10^{-3}$. Thus, there is another possibility that HFLS3 could 
be a pure star-forming source more compact than those of local pure star-forming LIRGs.

\subsection{Nature of Himiko}
\label{sec:nature}

In considering the origin of Himiko's extreme star formation rate and extensive Ly$\alpha$ halo,
it is convenient to return to the various explanations originally proposed by 
\citet{ouchi2009a} on the basis of the Subaru, UKIDSS, and shallow {\sl Spitzer} data available
at the time, taking into account the progress achieved with our new deep ALMA, {\sl HST}, and {\sl Spitzer}
data.

\smallskip
{\bf (i) A Gravitationally-Lensed Source}:

\cite{ouchi2009a} discounted this possibility on the basis of the resolved
kinematics of the extended $Ly\alpha$ halo. In Section \ref{sec:hst_observations}, we 
have strengthened the objections to this hypothesis since our {\sl HST} data reveal 
three $L^*$ sources whose morphological asymmetries are not consistent with
gravitational lensing. Moreover, we can find no potential foreground lens in the
vicinity of Himiko \citep{ouchi2009a}. Such a lens would have to be one of the three
clumps revealed in the {\sl HST} images, each of which has a $0.9\mu$m-continuum break
and blue UV continuum consistent with being physically associated at $z\simeq$6.6.
Our deep IRAC data also show no potential lensing sources near Himiko,
suggesting that there are no lensing objects with very red color,
such as dusty starbursts at intermediate redshifts, invisible in the optical and NIR bands.
Thus, we conclude that Himiko is not a gravitational lensed system.

\smallskip
{\bf (ii) Halo gas ionized by a hidden AGN}:

Our {\sl HST} images do not reveal any obvious point source that could represent
an active nucleus (Figure \ref{fig:HimikoHstSpitzerSubaru_images}). 
Moreover,
as noted by \citet{ouchi2009a}, Himiko is undetected at X-ray wavelengths
of $0.5-2$ keV down to $6\times 10^{-16}$ erg s$^{-1}$ cm$^{-2}$, and
the Keck optical spectrum does not reveal any high ionization features such 
as {\sc Nv}. 
Recently, very deep VLT/X-Shooter NIR spectroscopy has found no high ionization lines 
including {\sc Civ}$\lambda 1549$,
as would be expected for a hard ionizing AGN source
(Zabl et al. 2013 submitted to MNRAS).
Finally, radiative transfer simulations by \citet{baek2013} show
that the Ly$\alpha$ line profile and surface brightness of Himiko is inconsistent with 
heating from either a Compton-thick or Compton-thin AGN. Thus, we conclude the 
Ly$\alpha$ halo is unlikely to be heated by an active galactic nucleus.

\smallskip
{\bf (iii) Clouds of {\sc Hii} regions in a single virialized galaxy}:

\citet{ouchi2009a} discussed the possibility that Himiko could
be a single virialized system. Since the new
{\sl HST} data reveals three distinct UV luminous clumps each comparable
to the characteristic luminosity $L^*$, we consider
that Himiko is unlikely to be a single virialized system 
Although there are many
reports of disk galaxies with prominent clumps 
at $z\sim 2-3$ (e.g. \citealt{genzel2011}), the absence of
stellar disk (Figure \ref{fig:HimikoHstSpitzerSubaru_images}) 
distinguishes Himiko from a single galaxy with clumpy structures
such found at the lower redshifts.

\smallskip
{\bf (iv) Cold gas accretion onto a massive dark halo 
producing a central starburst}:

Some theoretical studies have suggested that cold gas can efficiently
penetrate into the central regions of a dark halo if that halo 
is more massive than 
the shock-heating scale of 
$\sim 4-7\times 10^{11}M_\odot$
at $z>4$ 
\citep{dekel2009,ocvirk2008}. 
Given Himiko's stellar mass
($1.5-3 \times 10^{10}M_\odot$, Section \ref{sec:stellar_population}) and 
little or no evolution in the ratio of stellar mass, $M_*$, to halo mass inferred
for $0<z<$1 ($M_*/M_{\rm DH}\lesssim 0.05$, \citet{leauthaud2012}), 
we expect a halo mass of $M_{\rm DH}\gtrsim 3-6\times 10^{11} M_\odot$. 
Abundance matching considerations support this estimate. \citet{ouchi2009a} 
calculated there should be at least one halo of mass $10^{12}M_\odot$ 
in the survey volume of $8\times 10^{5}$ comoving Mpc$^3$ 
where Himiko was found (see also \citealt{behroozi2012}). 

Although Himiko's halo mass does likely lie in the range
where cold accretion could be possible, we note that some 
recent simulations have cast doubt on the efficiency of this mode of 
assembly \citep{nelson2013,vogelsberger2013}.

\smallskip
{\bf (v) Outflowing gas excited by shocks or UV radiation from starbursts 
and/or mergers}:

The extensive Ly$\alpha$ nebula may be powered by star formation itself, but
the gas could also be shock heated by strong outflows driven by 
multiple supernova explosions in an intensive starburst \citep{mori2004}.
Figure \ref{fig:BetaLya} shows the relation between the UV slope $\beta$ 
which characterizes the stellar population and the Ly$\alpha$ equivalent width $EW_0$. 
This shows that the photoionization models of \citet{raiter2010}
whereby Ly$\alpha$ photons are scattered by the ISM and circum-galactic medium (CGM) 
can explain the properties of Himiko, notwithstanding the uncertainties in $\beta$. 
The success of this model depends, of course, on the escape fraction of ionizing
photons which should be moderately low ($<$50\%) so that scattering is
effective. However, the conclusion is robust even if we adopt a moderate 
dust extinction of $E(B-V)_s=0.15$ (Section \ref{sec:stellar_population}).

Thus, within the uncertainties, the amount of star formation observed is
sufficient to power the extended Ly$\alpha$ nebula; outflow and shocks are not required.
This simple photoionization scenario is consistent with the negligible hidden 
star-formation suggested by the weak dust and carbon emission
from our ALMA observations.
As discussed in \citet{ouchi2009a}, the FWHM of the Ly$\alpha$ line is only 
$v_{\rm FWHM}=251\pm 21$ km s$^{-1}$, further indicating that powerful
outflows are not present.

\smallskip
{\bf (vi) Merging bright galaxies}:

Although not a separate hypothesis from (v) above, we can ask what triggers
the intense star formation that likely powers the extended nebula.
In Figure \ref{fig:HimikoHstSpitzerSubaru_images}, we have identified three 
$L^*$ clumps which are highly suggestive of a rare triple merger.
As \citet{ouchi2009a} reported, Himiko presents a small
velocity offset of Ly$\alpha$ emission 
across the nebula ($\Delta v=60$km s$^{-1}$) with a narrow line width 
($v_{\rm FWHM} =251$ km s$^{-1}$; See Figure 7 of Ouchi et al. 2009a for
the Ly$\alpha$ line velocities that are measured on the slit position
shown with the red box in Figure 1 of Ouchi et al. 2009a).
Thus, the merger would have to
be largely confined to the direction perpendicular to the line of
sight.

Although a triple major merger is a rare event, 
our data suggest that the explanation is the most plausible. 
Recent numerical simulations predict that some extended Ly$\alpha$ sources
originate in mergers \citep{yajima2012b}. One interesting feature of Himiko 
is that the brightest portion of the Ly$\alpha$ nebula does not coincide
with the geometric center of the three clumps, but is located between 
the blue Ly$\alpha$ clump A and Clump B. Given the discussion in (v)
above, this indicates that Ly$\alpha$ photons are mainly produced by Clump A,
a very young and metal poor component.

\section{Summary}
\label{sec:summary}

We have taken deep ALMA and {\sl HST}/WFC3-IR data and supplementary
{\sl Spitzer} SEDS photometry for the remarkably luminous star-forming galaxy, Himiko, 
at $z$=6.595 which has a extended Ly$\alpha$ nebula.
at $z\sim 7$.
Following the original discovery \citep{ouchi2009a},these new data provide 
valuable insight into its physical properties and thereby offer an unique
perspective on how the earliest massive galaxies formed. We summarize our conclusions
as follows:

1. The 1.2mm dust continuum flux from this star-forming galaxy is very weak, $<52\mu$Jy, 
and comparable with or weaker than that observed for local dwarf irregulars with much
lower star formation rates.

2. We find a surprisingly stringent upper limit to the flux of the {\sc [Cii]} $158\mu$m line, 
$L_{\rm [CII]} < 5.4\times 10^{7} L_\odot$, 
placing it a factor $\simeq$30 below expectations
based on the scaling relation established between $L_{\rm [CII]}$ and star formation
rate for lower redshift galaxies. This indicates a very metal poor system and may
imply the [C II] line will be a poor diagnostic of early $z>$7 galaxies.

3. Our deeper {\sl HST}$+${\sl Spitzer} photometry allows us to considerably refine the
stellar population properties of Himiko. Using models with and without nebular lines,
we infer a stellar mass of $1.5-3\times 10^{10}M_\odot$ and a star formation rate
of $\simeq 100\pm 2 M_\odot$ yr$^{-1}$, comparable with the properties of 
luminous LBGs at $z\sim 3$.

4. Our {\sl HST} image has revealed three $L^*$ galaxy clumps which, together
our earlier kinematic constraints, suggests a rare triple merger. One clump 
reveals intense Ly$\alpha$ emission and an extremely blue color continuum
of $\beta = -2.84\pm 0.32$ suggestive of metal-poor star-formation and an
age of less than 200 Myr.

5. From these properties,we conclude we are witnessing intense star formation
induced by this triple merger and that the associated photoionizing radiation 
is sufficient to power the extensive Ly$\alpha$ nebula.

Although a rare object, Himiko has offered the first coherent view of  
how the most massive galaxies started forming at a time close to the
end of cosmic reionization at $z\sim 7$.

\acknowledgments
We are grateful to 
Pratika Dayal,
Andrea Ferrara,
Rob Kennicutt,
Kyoungsoo Lee,
Masao Mori,
Dominik Riechers,
Dimitra Rigopoulou,
Daniel Schaerer,
Masayuki Umemura,
Fabian Walter,
and
Chris Willott
for their useful comments and discussions.
We thank the ALMA observatory and {\sl HST} support
staff for their invaluable help 
that made these pioneering observations possible.
The {\sl HST} reduction was supported by a NASA STScI grant
GO 12265. 
This work was supported by World Premier International 
Research Center Initiative (WPI Initiative), MEXT, 
Japan, and KAKENHI (23244025) Grant-in-Aid for Scientific Research
(A) through Japan Society for the Promotion of Science (JSPS).
This paper makes use of the following ALMA data: ADS/JAO.ALMA\#2011.0.00115.S. 
ALMA is a partnership of ESO (representing its member states), NSF (USA) and 
NINS (Japan), together with NRC (Canada) and NSC and ASIAA (Taiwan), 
in cooperation with the Republic of Chile. The Joint ALMA Observatory is 
operated by ESO, AUI/NRAO and NAOJ.
This work is based on observations made with the NASA/ESA Hubble Space Telescope, 
obtained at the Space Telescope Science Institute, which is operated 
by the Association of Universities for Research in Astronomy, Inc., 
under NASA contract NAS 5-26555. These observations are associated 
with program \#12265. 
Support for program \#12265 was provided by NASA through a grant 
from the Space Telescope Science Institute, which 
is operated by the Association of Universities for Research in Astronomy, 
Inc., under NASA contract NAS 5-26555.
This work is based in part on observations made with the {\sl Spitzer} Space Telescope, 
which is operated by the Jet Propulsion Laboratory, 
California Institute of Technology under a contract with NASA. 
Support for this work was provided by NASA through an award issued by JPL/Caltech.

{\it Facilities:} \facility{ALMA (Band6) {\sl HST} (WFC3-IR) {\sl Spitzer} (IRAC)}

\clearpage

\clearpage


\begin{thebibliography}{}
\bibitem[Ashby et al.(2013)]{ashby2013} Ashby, M.~L.~N., Willner, S.~P., Fazio, G.~G., et al.\ 2013, \apj, 769, 80 
\bibitem[Baek \& Ferrara(2013)]{baek2013} Baek, S., \& Ferrara, A.\ 2013, \mnras, L77 
\bibitem[Behroozi et al.(2012)]{behroozi2012} Behroozi, P.~S., Wechsler, R.~H., \& Conroy, C.\ 2012, arXiv:1207.6105 
\bibitem[Vieira et al.(2013)]{vieira2013} Vieira, J.~D., Marrone, D.~P., Chapman, S.~C., et al.\ 2013, \nat, 495, 344 
\bibitem[Bouwens et al.(2010a)]{bouwens2010a} Bouwens, R.~J., et al.\ 2010, \apjl, 709, L133 
\bibitem[Bouwens et al.(2010b)]{bouwens2010b} Bouwens, R.~J., et al.\ 2010, \apjl, 708, L69 
\bibitem[Bouwens et al.(2011)]{bouwens2011} Bouwens, R.~J., Illingworth, G.~D., Oesch, P.~A., et al.\ 2011, \apj, 737, 90 
\bibitem[Bouwens et al.(2012)]{bouwens2012} Bouwens, R.~J., Illingworth, G.~D., Oesch, P.~A., et al.\ 2012, \apj, 754, 83 
\bibitem[Bruzual \& Charlot(2003)]{bruzual2003} Bruzual, G., \& Charlot, S.\ 2003, \mnras, 344, 1000 
\bibitem[Buat \& Xu(1996)]{buat1996} Buat, V., \& Xu, C.\ 1996, \aap, 306, 61 
\bibitem[Calzetti et al.(2000)]{calzetti2000} Calzetti, D., Armus, L., Bohlin, R.~C., Kinney, A.~L., Koornneef, J., \& Storchi-Bergmann, T.\ 2000, \apj, 533, 682
\bibitem[Capak et al.(2011)]{capak2011} Capak, P.~L., Riechers, D., Scoville, N.~Z., et al.\ 2011, \nat, 470, 233 
\bibitem[Castellano et al.(2010)]{castellano2010} Castellano, M., et al.\ 2010, \aap, 511, A20 
\bibitem[Chapman et al.(2005)]{chapman2005} Chapman, S.~C., Blain, A.~W., Smail, I., \& Ivison, R.~J.\ 2005, \apj, 622, 772 
\bibitem[Coppin et al.(2010)]{coppin2010} Coppin, K.~E.~K., Chapman, S.~C., Smail, I., et al.\ 2010, \mnras, 407, L103 
\bibitem[da Cunha et al.(2013)]{dacunha2013} da Cunha, E., Groves, B., Walter, F., et al.\ 2013, \apj, 766, 13 
\bibitem[Dale et al.(2007)]{dale2007} Dale, D.~A., Gil de Paz, A., Gordon, K.~D., et al.\ 2007, \apj, 655, 863 
\bibitem[Dayal et al.(2010)]{dayal2010} Dayal, P., Maselli, A., \& Ferrara, A.\ 2010, arXiv:1002.0839 
\bibitem[Dekel et al.(2009)]{dekel2009} Dekel, A., Birnboim, Y., Engel, G., et al.\ 2009, \nat, 457, 451 
\bibitem[de Looze et al.(2011)]{delooze2011} de Looze, I., Baes, M., Bendo, G.~J., Cortese, L., \& Fritz, J.\ 2011, \mnras, 416, 2712 
\bibitem[De Looze(2012)]{delooze2012} De Looze, I.\ 2012, Ph.D.~Thesis,  
\bibitem[Diaz-Santos et al.(2013)]{diazsantos2013} Diaz-Santos, T., Armus, L., Charmandaris, V., et al.\ 2013, arXiv:1307.2635 
\bibitem[Dunlop et al.(2012)]{dunlop2012} Dunlop, J.~S., McLure, R.~J., Robertson, B.~E., et al.\ 2012, \mnras, 420, 901 
\bibitem[Dunlop et al.(2013)]{dunlop2013} Dunlop, J.~S., Rogers, A.~B., McLure, R.~J., et al.\ 2012, arXiv:1212.0860 
\bibitem[Eales et al.(1989)]{eales1989} Eales, S.~A., Wynn-Williams, C.~G., \& Duncan, W.~D.\ 1989, \apj, 339, 859 
\bibitem[Ellis et al.(2013)]{ellis2013} Ellis, R.~S., McLure, R.~J., Dunlop, J.~S., et al.\ 2013, \apjl, 763, L7 
\bibitem[Finkelstein et al.(2010)]{finkelstein2010} Finkelstein, S.~L., Papovich, C., Giavalisco, M., et al.\ 2010, \apj, 719, 1250 
\bibitem[Finlator et al.(2006)]{finlator2006} Finlator, K., Dav{\'e}, R., Papovich, C., \& Hernquist, L.\ 2006, \apj, 639, 672 
\bibitem[Genzel et al.(2011)]{genzel2011} Genzel, R., Newman, S., Jones, T., et al.\ 2011, \apj, 733, 101 
\bibitem[Graci{\'a}-Carpio et al.(2011)]{gracia-carpio2011} Graci{\'a}-Carpio, J., Sturm, E., Hailey-Dunsheath, S., et al.\ 2011, \apjl, 728, L7 
\bibitem[Hayes et al.(2013)]{hayes2013} Hayes, M., {\"O}stlin, G., Schaerer, D., et al.\ 2013, \apjl, 765, L27 
\bibitem[Holland et al.(1999)]{holland1999} Holland, W.~S., Robson, E.~I., Gear, W.~K., et al.\ 1999, \mnras, 303, 659 
\bibitem[Iono et al.(2006)]{iono2006} Iono, D., Yun, M.~S., Elvis, M., et al.\ 2006, \apjl, 645, L97 
\bibitem[Ivison et al.(2010)]{ivison2010} Ivison, R.~J., Swinbank, A.~M., Swinyard, B., et al.\ 2010, \aap, 518, L35 
\bibitem[Kanekar et al.(2013)]{kanekar2013} Kanekar, N., Wagg, J., Ram Chary, R., \& Carilli, C.\ 2013, arXiv:1305.6469 
\bibitem[Kennicutt(1998)]{kennicutt1998} Kennicutt, R.~C., Jr.\ 1998, \araa, 36, 189 
\bibitem[Klaas et al.(1997)]{klaas1997} Klaas, U., Haas, M., Heinrichsen, I., \& Schulz, B.\ 1997, \aap, 325, L21 
\bibitem[Labb{\'e} et al.(2010)]{labbe2010} Labb{\'e}, I., Gonz{\'a}lez, V., Bouwens, R.~J., et al.\ 2010, \apjl, 716, L103 
\bibitem[Leauthaud et al.(2012)]{leauthaud2012} Leauthaud, A., Tinker, J., Bundy, K., et al.\ 2012, \apj, 744, 159 
\bibitem[Lee et al.(2012)]{lee2012} Lee, K.-S., Alberts, S., Atlee, D., et al.\ 2012, \apjl, 758, L31 
\bibitem[Maiolino et al.(2005)]{maiolino2005} Maiolino, R., Cox, P., Caselli, P., et al.\ 2005, \aap, 440, L51
\bibitem[Maiolino et al.(2009)]{maiolino2009} Maiolino, R., Caselli, P., Nagao, T., et al.\ 2009, \aap, 500, L1 
\bibitem[Malhotra et al.(2001)]{malhotra2001} Malhotra, S., Kaufman, M.~J., Hollenbach, D., et al.\ 2001, \apj, 561, 766 
\bibitem[Marsden et al.(2005)]{marsden2005} Marsden, G., Borys, C., Chapman, S.~C., Halpern, M., \& Scott, D.\ 2005, \mnras, 359, 43 
\bibitem[McLure et al.(2010)]{mclure2010} McLure, R.~J., Dunlop, J.~S., Cirasuolo, M., Koekemoer, A.~M., Sabbi, E., Stark, D.~P., Targett, T.~A., \& Ellis, R.~S.\ 2010, \mnras, 403, 960 
\bibitem[McLure et al.(2012)]{mclure2012} McLure, R.~J., Dunlop, J.~S., Bowler, R.~A.~A., et al.\ 2012, arXiv:1212.5222 
\bibitem[Meurer et al.(1999)]{meurer1999} Meurer, G.~R., Heckman, T.~M., \& Calzetti, D.\ 1999, \apj, 521, 64 
\bibitem[Mori et al.(2004)]{mori2004} Mori, M., Umemura, M., \& Ferrara, A.\ 2004, \apjl, 613, L97 
\bibitem[Nelson et al.(2013)]{nelson2013} Nelson, D., Vogelsberger, M., Genel, S., et al.\ 2013, \mnras, 429, 3353 
\bibitem[Ocvirk et al.(2008)]{ocvirk2008} Ocvirk, P., Pichon, C., \& Teyssier, R.\ 2008, \mnras, 390, 1326 
\bibitem[Oesch et al.(2010)]{oesch2010} Oesch, P.~A., et al.\ 2010, \apjl, 709, L16 
\bibitem[Ono et al.(2010a)]{ono2010a} Ono, Y., et al.\ 2010, \mnras, 402, 1580 
\bibitem[Ono et al.(2010b)]{ono2010b} Ono, Y., Ouchi, M., Shimasaku, K., et al.\ 2010, \apj, 724, 1524 
\bibitem[Ono et al.(2012)]{ono2012} Ono, Y., Ouchi, M., Mobasher, B., et al.\ 2012, \apj, 744, 83 
\bibitem[Ono et al.(2013)]{ono2013} Ono, Y., Ouchi, M., Curtis-Lake, E., et al.\ 2012, arXiv:1212.3869 
\bibitem[Ouchi et al.(1999)]{ouchi1999} Ouchi, M., Yamada, T., Kawai, H., \& Ohta, K.\ 1999, \apjl, 517, L19 
\bibitem[Ouchi et al.(2004a)]{ouchi2004a} Ouchi, M., et al.\ 2004a, \apj, 611, 660 
\bibitem[Ouchi et al.(2009a)]{ouchi2009a} Ouchi, M., et al.\ 2009a, \apj, 696, 1164 
\bibitem[Ouchi et al.(2009b)]{ouchi2009b} Ouchi, M., et al.\ 2009b, \apj, 706, 1136 
\bibitem[Ouchi et al.(2010)]{ouchi2010} Ouchi, M., Shimasaku, K., Furusawa, H., et al.\ 2010, \apj, 723, 869 
\bibitem[Pety et al.(2004)]{pety2004} Pety, J., Beelen, A., Cox, P., et al.\ 2004, \aap, 428, L21 
\bibitem[Raiter et al.(2010)]{raiter2010} Raiter, A., Schaerer, D., \& Fosbury, R.~A.~E.\ 2010, \aap, 523, A64 
\bibitem[Reddy \& Steidel(2009)]{reddy2009} Reddy, N.~A., \& Steidel, C.~C.\ 2009, \apj, 692, 778 
\bibitem[Reddy et al.(2010)]{reddy2010} Reddy, N.~A., Erb, D.~K., Pettini, M., Steidel, C.~C., \& Shapley, A.~E.\ 2010, \apj, 712, 1070 
\bibitem[Riechers et al.(2010)]{riechers2010} Riechers, D.~A., Capak, P.~L., Carilli, C.~L., et al.\ 2010, \apjl, 720, L131 
\bibitem[Riechers et al.(2013)]{riechers2013} Riechers, D.~A., Bradford, C.~M., Clements, D.~L., et al.\ 2013, arXiv:1304.4256 
\bibitem[Robertson et al.(2013)]{robertson2013} Robertson, B.~E., Furlanetto, S.~R., Schneider, E., et al.\ 2013, \apj, 768, 71 
\bibitem[Salpeter(1955)]{salpeter1955} Salpeter, E.~E.\ 1955, \apj, 121, 161 
\bibitem[Sargsyan et al.(2012)]{sargsyan2012} Sargsyan, L., Lebouteiller, V., Weedman, D., et al.\ 2012, \apj, 755, 171 
\bibitem[Sauvage \& Thuan(1994)]{sauvage1994} Sauvage, M., \& Thuan, T.~X.\ 1994, \apj, 429, 153 
\bibitem[Schaerer \& de Barros(2009)]{schaerer2009} Schaerer, D., \& de Barros, S.\ 2009, \aap, 502, 423 
\bibitem[Schaerer \& de Barros(2010)]{schaerer2010} Schaerer, D., \& de Barros, S.\ 2010, \aap, 515, A73 
\bibitem[Schenker et al.(2012)]{schenker2012} Schenker, M.~A., Robertson, B.~E., Ellis, R.~S., et al.\ 2012, arXiv:1212.4819 
\bibitem[Silva et al.(1998)]{silva1998} Silva, L., Granato, G.~L., Bressan, A., \& Danese, L.\ 1998, \apj, 509, 103 
\bibitem[Stacey et al.(2010)]{stacey2010} Stacey, G.~J., Hailey-Dunsheath, S., Ferkinhoff, C., et al.\ 2010, \apj, 724, 957 
\bibitem[Storchi-Bergmann et al.(1994)]{storchi-bergmann1994} Storchi-Bergmann, T., Calzetti, D., \& Kinney, A.~L.\ 1994, \apj, 429, 572 
\bibitem[Vallini et al.(2013)]{vallini2013} Vallini, L., Gallerani, S., Ferrara, A., \& Baek, S.\ 2013, \mnras, 1486 
\bibitem[Vogelsberger et al.(2013)]{vogelsberger2013} Vogelsberger, M., Genel, S., Sijacki, D., et al.\ 2013, arXiv:1305.2913 
\bibitem[Venemans et al.(2012)]{venemans2012} Venemans, B.~P., McMahon, R.~G., Walter, F., et al.\ 2012, \apjl, 751, L25 
\bibitem[Walter et al.(2009)]{walter2009} Walter, F., Riechers, D., Cox, P., et al.\ 2009, \nat, 457, 699 
\bibitem[Walter et al.(2012)]{walter2012} Walter, F., Decarli, R., Carilli, C., et al.\ 2012, \apj, 752, 93 
\bibitem[Wang et al.(2013)]{wang2013} Wang, R., Wagg, J., Carilli, C.~L., et al.\ 2013, arXiv:1302.4154 
\bibitem[Willott et al.(2013)]{willott2013} Willott, C.~J., Omont, A., \& Bergeron, J.\ 2013, \apj, 770, 13 
\bibitem[Yajima et al.(2012a)]{yajima2012a} Yajima, H., Umemura, M., \& Mori, M.\ 2012, \mnras, 420, 3381 
\bibitem[Yajima et al.(2012b)]{yajima2012b} Yajima, H., Li, Y., \& Zhu, Q.\ 2012, arXiv:1210.6440 
\bibitem[Yan et al.(2005)]{yan2005} Yan, H., Dickinson, M., Stern, D., et al.\ 2005, \apj, 634, 109 
\bibitem[Young et al.(1989)]{young1989} Young, J.~S., Xie, S., Kenney, J.~D.~P., \& Rice, W.~L.\ 1989, \apjs, 70, 699 
\end{thebibliography}
\end{document}